# Differentiation of Vesta: Implications for a shallow magma ocean

Wladimir Neumann[*1], Doris Breuer[1] and Tilman Spohn[1,2]

[1]Institute of Planetary Research, German Aerospace Center (DLR)

Rutherfordstraße 2, 12489 Berlin, Germany

[2]Institute of Planetology, Westfälische Wilhelm-University Münster,

Wilhelm-Klemm-Str. 10, 48149 Münster, Germany

E-Mail: [*]Wladimir.Neumann@dlr.de, Doris.Breuer@dlr.de, Tilman.Spohn@dlr.de

*Corresponding Author: Wladimir Neumann, Tel. +493067055395, Fax. +493067055303

25  **Proposed Running Head:** Differentiation of Vesta: Implications for a shallow magma ocean

26  **Editorial Correspondence to:**

27  Wladimir Neumann

28  Institute of Planetary Research

29  German Aerospace Center (DLR)

30  Rutherfordstraße 2

31  12489 Berlin

32  Germany

33  Phone: +493067055395

34  Fax: +493067055303

35  E-Mail address: Wladimir.Neumann@dlr.de

# Abstract

The Dawn mission confirms earlier predictions that the asteroid 4 Vesta is differentiated with an iron-rich core, a silicate mantle and a basaltic crust, and supports the conjecture of Vesta being the parent body of the HED meteorites. To better understand its early evolution, we perform numerical calculations of the thermo-chemical evolution adopting new data obtained by the Dawn mission such as mass, bulk density and size of the asteroid.

We have expanded the thermo-chemical evolution model of Neumann et al. (2012) that includes accretion, compaction, melting and the associated changes of the material properties and the partitioning of incompatible elements such as the radioactive heat sources, advective heat transport, and differentiation by porous flow, to further consider convection and the associated effective cooling in a potential magma ocean. Depending on the melt fraction, the heat transport by melt segregation is modelled either by assuming melt flow in a porous medium or by simulating vigorous convection and heat flux of a magma ocean with a high effective thermal conductivity.

Our results show that partitioning of $^{26}$Al and its transport with the silicate melt is crucial for the formation of a global and deep magma ocean. Due to the enrichment of $^{26}$Al in the liquid phase and its accumulation in the sub-surface (for formation times $t_0$<1.5 Ma), a thin shallow magma ocean with a thickness of 1 to a few tens of km forms above which a basaltic crust forms – its thickness depends on the viscosity of silicate melt. The lifetime of the shallow magma ocean is $O(10^4)$-$O(10^6)$ years and convection in this layer is accompanied by the extrusion of $^{26}$Al at the surface. The interior differentiates from the outside inward with a mantle that is depleted in $^{26}$Al and core formation is completed within ~0.3 Ma. The lower mantle experiences a maximal melt fraction of 45% suggesting a harzburgitic to dunitic composition. Our results support the formation of eucrites by the extrusion of early partial melt and cumulative eucrites and diogenites may form from the crystallizing shallow magma ocean. Silicate melt is present in the mantle for up to 150 Ma, and convection in a crystallizing core proceeds for approximately 100 Ma, supporting the idea of an early magnetic field to explain the remnant magnetisation observed in some HED meteorites.

**Key Words:** Asteroids, dynamics; Interiors; Meteorites; Planetary formation; Planetesimals.

## 1. Introduction

The large asteroid 4 Vesta, located in the inner belt at a mean heliocentric distance of 2.36 AU, is the second-most-massive body in the asteroid belt. Ground-based observations indicate a dry, basaltic surface composition suggesting that melt segregation and core-mantle differentiation must have taken place and that the body has been resurfaced by basaltic lava flows (e.g., Zellner et al. 1997). The Dawn mission recently orbiting Vesta confirms this observation and the notion that this asteroid is the parent body of the HED (howardite, eucrite and diogenite) meteorites (De Sanctis et al. 2012, Russell et al. 2013). HED's consist mainly of noncumulative eucrites (pigeonite-plagioclase basalts) and orthopyroxene-rich diogenites. Howardites are impact breccias, composed predominantly of eucrite and diogenite clasts. The igneous lithology is completed with more rare rock types that include cumulative eucrites and olivine-bearing diogenites. This lithology has been used to constrain the interior structure. Vesta is believed to have a layered structure consisting of at least a metallic core, a rocky olivine-rich mantle and a crust consisting mainly of an upper basaltic (eucritic) and a lower orthopyroxene-rich layer (e.g., Delany, 1995; Ruzicka et al., 1997; Righter and Drake, 1997; Mandler and Elkins-Tanton, 2013). The details in the interior structure vary between the models and depend among others on the assumption of the bulk composition of Vesta. A eucritic upper crust and orthopyroxene-rich lower crust have been also identified by Dawn in particular in the Rheasilvia basin (De Sanctis et al. 2012, Prettyman et al. 2012). The huge basin allows the identification of material that was formed down to a depth of 30-45 km and has been excavated or exposed by an impact (Jutzi and Asphaug, 2011; Ivanov and Melosh, 2013). Interestingly, olivine-rich and thus mantle material has not been found in a significant amount, i.e. not above the detection limit of <25%, suggesting a crustal thickness of at least 30-45 km (McSween et al. 2013). This finding may also explain the lack of olivine–rich samples in the HED collection. It should be noted though that several olivine-rich terrains were actually detected by Dawn in the northern hemisphere (De Sanctis et al. 2013, Ruesch et al. 2013).

Two possible differentiation scenarios have been associated with the formation of the basaltic achondrites (i.e., eucrites and diogenites). The first scenario suggests that eucrites and diogenites originated from the partial melting of the silicates (e.g. Stolper, 1975, 1977; Jones et al. 1984; Jones et al., 1996) with the extraction of basaltic (euritic) magma leaving behind a harzburgite, orthopyroxenite or dunite residual depending on the degree of partial melting. The other scenario favored by geochemical arguments suggests achondrites being cumulates formed by magma fractionation. In the latter scenario the diogenites could have crystallized either in a magma ocean (Ikeda and Takeda, 1985; Righter and Drake, 1997; Ruzicka et al., 1997; Takeda, 1997; Warren, 1997; Drake, 2001; Greenwood et al., 2005; Schiller et al., 2011) or in multiple, smaller magma chambers (Shearer et al., 1997; Barrat et al., 2008; Beck and McSween, 2010; Mandler and Elkins-Tanton, 2013). Eucrites are then products from the magmas that had earlier crystallized diogenites.

Eucrites exhibit siderophile depletions that indicate the formation of an iron-rich core prior to their crystallization and within ~1 to 4 Ma of the beginning of the solar system (Palme and Rammensee 1981; Righter and Drake 1997, Kleine et al., 2009). Assuming core densities between 7000 and 8000 kg m$^{-3}$, the core radius of Vesta is estimated to lie between 105 and 114 km (Raymond et al. 2012, Russell et al. 2012), approximately half of the asteroid's radius. Some HEDs (e.g. Millbillillie and Allan Hills A81001), furthermore, show a remnant magnetisation suggesting a formerly active dynamo in a liquid metallic core (Fu et al. 2012a,b).

To understand Vesta's thermal and geological evolution, several numerical and experimental studies (e.g. Righter and Drake 1997, Ghosh and McSween 1998, Drake 2001, Gupta and Sahijpal 2010) have been performed. Questions these studies want to answer are related to the timing of accretion and core formation and the history of volcanism, especially the formation of the basaltic crust. Related to the last issue is in particular the question about the origin of the HED meteorites that either originated from the partial melting of the silicates or are residual melts of a crystallizing and formerly convecting whole-mantle magma ocean. It should be noted though that in most models the formation of core and crust has not been modelled self-consistently; rather some specific scenario has been assumed and its consequences for the thermal evolution on Vesta have been studied.

Righter and Drake (1997) considered core formation and crystallisation of a cooling magma ocean using a numerical physicochemical model. They suggested that after cooling to a crystal fraction of 80%, residual melt percolated from the former extensive magma ocean to form eucrites at the surface and diogenites in shallow layers by further crystallisation. In their model complete crystallisation occurred within 20 Ma after the formation of Vesta.

Ghosh and McSween (1998) investigated the differentiation of Vesta by assuming instantaneous core formation in the temperature interval of 1213-1223 K (assumed solidus and liquidus temperature of Fe-FeS) and that HED meteorites are the product of 25% partial melting. To obtain such a scenario, they concluded that Vesta must have accreted at 2.85 Ma, differentiated at 4.58 Ma and formed a basaltic crust at 6.58 Ma relative to the formation of the CAIs. Furthermore, they suggested that the mantle remained hot for 100 Ma after its formation and that some near-surface layers may have remained undifferentiated.

The latter models suggesting a global magma ocean have in common that they neglect efficient cooling due to convection in the magma ocean and partitioning of $^{26}$Al into the silicate melt and of its migration towards the surface. So far, there have not been many attempts to model magma ocean convection in planetesimals,. A first model to study the influence of a convecting magma ocean that formed due to heating by $^{26}$Al in planetesimals with radii of less than 100 km was presented by Hevey and Sanders (2006). Convection and the associated heat transport have been simulated by increasing the thermal conductivity by three orders of magnitude upon reaching a melt fraction of 50% - at this rheologically critical melt fraction (RCMF) a strong decrease in the viscosity is expected resulting in vigorous convection and efficient heat transport (Marsh, 1988). As an important consequence, it has been shown that the excess radiogenic heating does not raise the temperature of the convecting interior and thus the degree of partial melting in the magma ocean but rather increases the extent of the molten zone.

Gupta and Sahijpal (2010) adopted this approach in their simulations for Vesta by varying the thermal diffusivity by three orders of magnitude between the melt fractions of 50% and 100% (thereby, the thermal conductivity was increased). They investigated two evolution paths, namely the formation of basaltic achondrites in Vesta via partial melting of silicates or as residual melts after crystallisation of a convecting magma ocean. They concluded that, depending on the formation time, melt extraction is possible between 0.15 and 6 Ma after the CAIs and that differentiation proceeds rapidly within $O(10^4)$ a. For the scenarios where accretion was completed within 2 Ma after the CAIs, a magma ocean formed, which does not crystallise completely for at least 6-10 Ma. We will later show that for magma oceans comprising a large part of the mantle, the convective heat flux is underestimated in the studies Hevey and Sanders (2006) and Gupta and Sahijpal (2010) in comparison to that of the soft turbulence regime that is valid when the viscosity is small and convection is extremely turbulent as in the case of magma oceans (Solomatov, 2007). In fact, as we point out further below, the effective thermal conductivity of the magma ocean is underestimated by up to three orders of magnitude.

The importance of equilibrium partitioning of $^{26}$Al into the silicate melt and of its migration toward the surface has been shown by Moskovitz and Gaidos (2011) for planetesimals smaller than Vesta. They concluded that migration of $^{26}$Al-enriched melt towards the surface inhibits a further increase of internal temperatures and partial melting. In their model, only subsequent heating by $^{60}$Fe generates melt fractions of over 50% completing the differentiation of the bodies of interest. Although Vesta was not considered in their study, the partitioning of $^{26}$Al, its extrusion to the surface and the associated depletion of the interior would have a strong impact on its thermal evolution.

As a general conclusion from the previous thermal models one can state that Vesta should have finished its accretion by ≈3 Ma after the CAIs due to the decrease of the radiogenic heating by $^{26}$Al and $^{60}$Fe with time. Whether the basaltic crust, i.e. eucrites and diogenites, did form after the core-mantle separation due to the crystallisation of a global magma ocean (residual melt origin) or prior to the core-mantle separation due to the extrusion of the first partial melts (partial melting origin) remains an open issue, although geochemical models suggest that core formation precedes crust formation (e.g. Kleine et al., 2009). Which scenario is more likely also depends on the competition between heating by short-lived isotopes and the depletion of these heat sources due to their partitioning into the melt and the migration velocity of this melt by porous flow. Only if the extrusion of melt containing $^{26}$Al is comparatively slow an asteroid-scale magma ocean may have formed, otherwise HEDs formed by the extrusion of early partial melts. In case of a magma ocean, its convective cooling and the associated thermal convection in the core could result in a core dynamo lasting for several 10 Ma if Vesta formed before ≈1.5 Ma after the CAIs (Elkins-Tanton et al. 2011) although Sterenborg and Crowley (2013) argue that only parent bodies larger than 500 km can produce a sufficient strong magnetic field longer than 10 Ma by thermal convection.

In the present study, we investigate the thermal evolution of Vesta by means of a comprehensive model that includes melting, associated changes of the material properties, differentiation via porous flow and the corresponding advective heat transport below the RCMF of 50% (Marsh, 1988), as well as convection and the associated effective cooling in a magma ocean and liquid core for melt fractions above the RCMF applying the soft-turbulence approach (Solomatov et al., 2007). Numerical calculations of the thermo-chemical evolution of Vesta are performed adopting the new data obtained by the Dawn mission (e. g. the mass, the bulk density and the dimensions, see Russell et al. 2012). In particular, we investigate the occurrence and the time scale of core-mantle differentiation, the differences between the models that consider partitioning of $^{26}$Al and those that neglect partitioning of $^{26}$Al, as well as the influence of the convecting magma ocean on the thermal evolution of Vesta. We conclude that due to the extrusion of $^{26}$Al towards the surface a thin sub-surface magma ocean forms prior to the completion of the core-mantle differentiation and that the depleted mantle remains below the RCMF.

## 2. Model description

The evolution of an asteroid begins with its accretion from the protoplanetary dust as a porous aggregate. Heating by radioactive decay combined with self-gravity leads to substantial compaction at sub-solidus temperatures. Provided strong heating, the metal phase and subsequently the silicate phase starts to melt upon reaching the respective solidus temperatures. The radioactive isotope $^{26}$Al partitions among other incompatible elements into the silicate melt. If melting is wide-spread enough, the asteroid can differentiate (at least partially) via porous flow into a metallic core and a silicate mantle. Differentiation is initiated by the iron melt sinking downward and silicate melt percolating towards the surface. The mantle and core regions, in which the partial melt increases above 50%, start convecting. Furthermore, convection starts only when the heat flow at the upper interface becomes super-adiabatic – this is in particular a limiting factor for convection in the core. The asteroid cools due to convection and after the melt fractions drop below 50% by conduction and melt heat transport to the sub-solidus temperatures. After cooling below the solidus, no further structural changes (differentiation) of the interior can be expected except by external influence such as impacts.

To model the above evolution path for Vesta, we have adopted the thermal evolution model from Neumann et al. (2012) which considers accretion, compaction, melting and differentiation by porous flow of ordinary chondritic planetesimals. For a detailed description

256  of the model we refer to Neumann et al. (2012) and Neumann et al. (2013). Next, we give a

257  brief summary of the important model features and describe in particular the additional

258  processes implemented for the current study as well as the parameters that are specific for

259  Vesta.

260  The rate of change of internal energy, including the consumption and release of latent heat

261  due to melting and solidification is computed from the conductive heat flow, the volumetric

262  heat production rate by radioactive decay, and the advective heat transport by melt

263  percolation:

$$\rho c_p \left(1 + x_{Fe} Ste_{Fe} + x_{Si} Ste_{Si}\right) \frac{\partial T}{\partial t} = \vec{\nabla}\left(k(\vec{x})\vec{\nabla}T\right) + Q(\vec{x},t) + \vec{\nabla}T\left(\Delta v_{Fe} + \Delta v_{Si}\right), \quad (1)$$

265  where $\rho$ is the density, $c_p$ is the specific heat capacity, $x_{Fe}$ and $x_{Si}$ are the current mass fractions

266  of metallic iron and silicates, respectively, $Ste_{Fe}$ and $Ste_{Si}$ are the Stefan numbers of metal and

267  silicates, respectively, $T$ is temperature, $t$ is time, $\vec{x}$ is the space variable, $k$ is the thermal

268  conductivity, $Q$ is the heat source density, and $\Delta v_{Fe}$ and $\Delta v_{Si}$ are the velocities of iron and

269  silicate melts relative to the matrix, respectively.

270  Because we consider porosity loss during the thermal evolution, the initial radius $R_p$ of the

271  body changes to the compacted radius $\bar{R}_p(t) = R_p \left(\frac{1-\phi_0}{1-\bar{\phi}(t)}\right)^{1/3}$, where $\bar{\phi}(t)$ is the average

272  porosity at time $t$, and $\phi_0$ is the uniform initial porosity. The average porosity $\bar{\phi}(t)$ and the

273  average density $\bar{\rho}(t)$ are computed using the equations $\bar{\phi}(t) = 1 - V_i/V_b$ and $\bar{\rho}(t) = M/V_b$ where

274  $V_i$ is the intrinsic volume (i.e., the volume of the solid material), $V_b$ is the bulk volume (i.e.,

275  the volume of the solid material plus the volume of the pores, computed by integrating the

276  porosities at the grid points of the discretisation over the body) and $M$ is the mass of Vesta.

277  The thermal conductivity, density, heat source density, radius, and some other parameters

278  change with time due to the loss of porosity and melting. The loss of the porosity $\phi$ is a

279  consequence of compaction by hot-pressing under the influence of temperature and pressure:

$$\frac{\partial \log(1-\phi(\vec{x}))}{\partial t} = A\sigma^{2/3}b^{-3}e^{-E/RT}, \qquad (2)$$

where $A=4.0\cdot 10^{-5}$ is a constant, $\sigma$ is the effective stress acting on a unit volume, $b$ is the initial grain size, $E$ is the activation energy, and $R$ is the universal gas constant.

Iron or silicate melt is generated linearly between the solidus and the liquidus temperatures of metal or silicates. In this temperature interval both components are present in both solid and liquid states. Consumption and release of the latent heat during the melting and solidification buffers temperature variations between the entirely solid and the entirely liquid state.

In terms of the flow in porous media theory, we relocate the intrinsically denser (than the bulk solid matrix) metal melt downwards, and the intrinsically lighter silicate melt towards the surface, according to Darcy's law. This approach is valid below the RCMF. In this case, the energy transport by the melts enters into the heat conservation equation via advection terms. Thereby, the associated grain growth due to Ostwald ripening is accounted for in the partially molten regions of the body resulting in faster percolation velocities with time.

Assuming equilibrium melting, the chemical partitioning of $^{26}$Al is described by

$$c_{Al}^{tot} = \chi_{Si}c_{Al}^{l} + (1-\chi_{Si})c_{Al}^{s}, \quad c_{Al}^{l} = \frac{c_{Al}^{tot}}{P+\chi_{Si}(1-P)}, \quad c_{Al}^{s} = \frac{c_{Al}^{tot}P}{P+\chi_{Si}(1-P)}, \qquad (5)$$

with the total concentration of $^{26}$Al $c_{Al}^{tot}$, the concentration in the solid and liquid phase $c_{Al}^{s}$ and $c_{Al}^{l}$, the melt fraction of silicates (relative to the overall amount of silicates) $\chi_{Si}$, and the partitioning coefficient $P$ (Best, 2002).

To calculate the initial composition of the primordial material we assume that Vesta formed from L-chondritic material. The minerals, their mass fractions and densities are taken from Yomogida and Matsui (1983) (see Table 1). The metal phase consists of free iron and nickel and of the troilite. The remaining minerals form the silicate phase. This composition is a reasonable choice considering Dawn's measurements of the density. First, a metal mass fraction of 0.14 in an L-chondritic composition results in a relative core radius of ≈0.43 if one assumes complete differentiation, which agrees with the core radius of Vesta suggested by Russell et al. (2012). Note that for an H-chondritic composition a larger core with a relative core radius of ≈0.49 would follow (cf. Neumann et al. 2012). Second, the average bulk density of 3456±34 kg m$^{-3}$ (Russell et al. 2012) lies slightly below a typical intrinsic density of an L-chondrite (see Yomogida and Matsui 1983), which is approximately 3600 kg m$^{-3}$ (cp. Table 1).

Heavy cratering of the surface suggests a porous upper mantle and crust. Russell et al. (2012) suggested a porosity of 5-6% in these layers, which means that the average porosity of the entire body is smaller. From $\rho_i = \rho_b (1-\phi)^{-1}$ where $\rho_i$ is the average intrinsic density and $\rho_b$ is the bulk density, the average porosity of Vesta $\phi$ can be calculated. For $\rho_b = 3490$ kg m$^{-3}$ which is the smallest upper bound we obtain $\phi = 0.03$ and for $\rho_b = 3456$ kg m$^{-3}$, $\phi = 0.04$ follows. This would mean that the average porosity of the mantle and the crust of Vesta is larger than 4%. The bulk density $\rho_{b0}$ of the primordial material during the accretion of Vesta is calculated by using a reasonable uniform initial porosity $\phi_0 = 0.4$ of the accreting material (see Henke et al. 2012, Neumann et al. 2012) and the equation $\rho_{b0} = (1-\phi_0)\rho_i$. For $\rho_i = 3600$ kg m$^{-3}$ we obtain $\rho_{b0} = 2160$ kg m$^{-3}$.

Taking the volume of Vesta as 7.497x10$^7$ m$^3$ (Russell et al. 2012) and assuming a negligible average porosity (in our most realistic calculations Vesta compacted almost completely; this is a result of our model runs) a radius of 262 km is obtained. We take this as a reference value which corresponds to a completely compacted body. For an appropriate choice of the average present-day porosity the value of 262.7±0.1 inferred by Russell et al. (2012) follows. Note though that a body with the axes of 572.6 by 557.2 by 446.4 km is considered here to be spherically symmetric. Our model combines the average (uniform) initial porosity of $\phi_0 = 0.4$ with the reference radius of $D$=262 km to obtain the initial radius of $R_p$=310 km.

328  For melt fractions above 50%, the settling velocity $v_s$ of the molten iron globules or solid iron

329  crystals in a silicate mush is computed via Stokes´ law

330 $$v_s = \frac{2}{9}\frac{\rho_g - \rho_m}{\eta_{l,Si}}gb^2, \tag{3}$$

331  where $\rho_g$ is the density of a metal globule (liquid iron), $\rho_m$ is the density of the silicate mush

332  (liquid silicate), $\eta_{l,Si}$ is the viscosity of the liquid silicates, $g$ is the gravitational acceleration

333  and $b$ is the radius of a metal globule (here a constant value of 0.01 m is adopted, see e.g.

334  Samuel, 2012 and Taylor, 1992 for the globule size). Strictly speaking, further dynamic

335  regimes (Newton and intermediate regimes, see Samuel, 2012) should be considered besides

336  the Stokes regime. Our estimates using parameters from Tables 1 and 2 indicate that the

337  velocities obtained in the Stokes, Newton or intermediate regimes are very similar and much

338  higher compared to the Darcy velocity obtained below the RCMF (the latter is orders of

339  magnitude smaller). The metal separation is thus almost instantaneous for the situation of high

340  melt fraction and we therefore consider only the Stokes regime.

341  Throughout the study the nebula temperature of $T_n=290$ K (this value corresponds to the

342  temperature in the Solar nebula at 2.36 AU, see Ghosh and McSween 1998) is used as the

343  initial temperature of the accreting material and also for the boundary condition. The radiation

344  boundary condition is adopted with the emissivity e=0.8 (Ghosh and McSween 1998).

345  For a body which formed during the first few million years after the formation of the Solar

346  system the short-lived isotopes $^{26}$Al and $^{60}$Fe are believed to be the most important heat

347  sources. Whereas most researchers suggest the "canonical" ratio $[^{26}Al/^{27}Al]_0=5\times10^{-5}$ at the

348  time of the formation of CAIs, there is still an uncertainty about the initial ratio $[^{60}Fe/^{56}Fe]_0$

349  (values between $10^{-8}$ and $10^{-6}$ have been inferred by numerous workers). For $^{26}$Al we use the

350  canonical ratio $[^{26}Al/^{27}Al]_0=5\times10^{-5}$ at the formation of the CAIs, and for $^{60}$Fe the value

351  $[^{60}Fe/^{56}Fe]_0=1.6\times10^{-6}$ reported by Cameron (1993). As this value is highly uncertain and

352  recent experiments suggest that previous measurements are likely biased and predict to high values of $[^{60}Fe/^{56}Fe]_0$ (Ogliore et al., 2011; Telus et al., 2012), we further study the influence of the amount of $^{60}Fe$ on the results.

Apart from the initial ratios above, the heat production rate depends on the abundance of Al and Fe in the bulk material. The latter are computed from the exact composition and fractions of metal and silicates (see further below). The heat sources density term $Q(r,t)$ in the energy balance equation (1) arises from the radiogenic decay of $^{26}Al$ and $^{60}Fe$. We write this term as

$$Q(r,t) = \rho \left( v_{Si} \left[ \frac{^{26}Al}{^{27}Al} \right] \frac{E_{Al}}{\tau_{Al}} e^{-\frac{t+t_0}{\tau_{Al}}} + \frac{x_{Fe}}{x_{Fe,0}} f_{Fe} \left[ \frac{^{60}Fe}{^{56}Fe} \right] \frac{E_{Fe}}{\tau_{Fe}} e^{-\frac{t+t_0}{\tau_{Fe}}} \right), \qquad (4)$$

where $v_{Si} = f_{Al} \, x_{Si}/x_{Si,0}$ (prior to silicate melting, partitioning and melt segregation), $x_{Fe,0}$ and $x_{Si,0}$ are the initial mass fractions of metal and silicates, $f_{Fe}$ and $f_{Al}$ is the number of iron or aluminium atoms per 1 kilogram of the primordial material, respectively, $E_i$ is the decay energy per atom, $\tau_i$ is the mean life, $[^*i/^*i]_0$ is the initial abundance of the isotope $^*i$, and $t_0$ is the delay time of accretion with respect to the formation time of the CAIs. The mass fraction ratios control the changes in the heat production due to the differentiation. The bulk Al mass fraction is calculated from that of $Al_2O_3$ in L-chondrites reported by Jarosevich (1990). The latter value of 0.0225 is multiplied with the ratio of the atomic mass of $^{27}Al$ (in g mol$^{-1}$) times two, and the sum of the atomic masses of $^{27}Al$ times two and $^{16}O$ times three, to obtain the value of 0.0119. This is slightly larger than the value of 0.0113 used by Ghosh and McSween (1998) for Vesta, because they assumed an H-chondritic composition and adopted the $Al_2O_3$ mass fraction of 0.0214 from Jarosevich (1990).

The mass fraction of the bulk iron is computed from the mass fraction of the free iron Fe and the mass fraction of FeS times the ratio of the atomic mass of Fe and the sum of the atomic masses of Fe and S. Thereupon, the (initial) abundances are computed using the formulas $f_{Al} = 10^3 m_{Al} N_A (ma_{26Al})^{-1}$ (before melting and partitioning) and $f_{Fe} = 10^3 m_{Fe} N_A (ma_{60Fe})^{-1}$, where $m_{Al}$ and $m_{Fe}$ are mass fraction of the elements, $ma_{26Al}$ and $ma_{60Fe}$ are the atomic masses of

377 the heat generating isotopes, and $N_A$ is the Avogadro number (for the resulting values see

378 Table 2). The factor $v_{Si}$ controls on the one hand the changes in the silicate mass fraction due

379 to the migration of the silicate melt as compared to the initial mass fraction. On the other

380 hand, this factor updates the changes in the concentration of $^{26}$Al in a given discretisation

381 shell, which arise from the mixing of migrated and present melt with different concentration

382 of $^{26}$Al and from the matrix compaction. Once melt segregation starts, $^{26}$Al particles

383 associated with the migrated silicate melt from a lower shell are added to those remaining in

384 the upper shell after the compaction of the mixed metal-silicate-matrix. The number of $^{26}$Al

385 particles in the lower shell is updated to account for the $^{26}$Al that left with the silicate melt and

386 the particles added due to the matrix compaction. Similarly, as iron melt moves from an upper

387 shell to a lower one and is being replaced by either a certain volume of the matrix or/and a

388 certain volume of the silicate melt, the number of $^{26}$Al particles is updated due to these

389 volume changes. We have neglected the heating by $^{60}$Fe which remains in the silicate phase

390 after core formation, its contribution to the heating of the mantle is, however, negligible.

391 As in the previous studies, we do not consider the solid-state convection for melt fractions

392 below 50%. In order to model the convection in a magma ocean, where more than 50% of the

393 material is molten, we adopt the method presented in Solomatov (2007) (see also Kraichnan

394 1962 and Siggia 1994) to compute the convective heat flux in the soft turbulence regime

395 (when the viscosity is small and convection is extremely turbulent, but the Rayleigh number is

396 still below approximately $10^{19}$). This approach has also been used for modelling of local

397 magma ponds on Mars (Golabek et al. 2011). From the convective heat flux the effective

398 thermal conductivity is calculated. The thermal conductivity in the magma ocean is

399 substituted by the effective thermal conductivity, which mimics the heat flux of a medium

400 with a low magma ocean viscosity:

401 $k_{eff} = F_{conv} L / \Delta T,$ (5)

where $F_{conv}$ is the convective heat flux, $L$ is the depth of the magma ocean and $\Delta T = T_p - T_s$, with the potential temperature $T_p$ and the temperature $T_s$ at the upper boundary of the magma ocean. In a silicate magma ocean this temperature corresponds to the silicate melt fraction of 50% and is equal to 1637 K. In a convecting core this is the temperature at the core-mantle boundary or at the upper boundary of the convecting part of the core. The latter is equal to or higher than the temperature of 1456 K which corresponds to the iron melt fraction of 50 %. In the following we use the index $i$ which is either *Fe* or *Si* to distinguish between a silicate magma ocean in the mantle and a convecting (part of the) core. The convective heat flux in the soft turbulence regime may be expressed as follows:

$$F_{conv,i} = 0.089 k_i (T_{p,i} - T_{s,i}) Ra_i^{1/3} / L_i, \qquad (6)$$

where $k_i$ is the thermal conductivity and $Ra_i$ is the Rayleigh number. The Rayleigh number is computed from

$$Ra_i = \frac{\alpha_i g_i (T_{p,i} - T_{s,i}) \rho_i^2 c_{p,i} L_i^3}{k_i \eta_{l,i}} \qquad (7)$$

where $\alpha_i$ is the thermal expansivity, $g_i$ is the surface gravity of the respective convecting layer, $\rho_i$ is the density, $c_{p,i}$ is the heat capacity, and $\eta_{l,i}$ is the viscosity of the melt.

Convection in the core is treated in a similar manner. Here, if the heat flux at the core-mantle boundary from the core into the mantle is positive and climbs above the adiabatic heat flux at the CMB, and the melt fraction in the core is above the RCMF, thermal convection is assumed to occur (Stevenson 2003). This is achieved by solving equations (5)-(7) using the parameters of the core.

The adiabatic heat flow at the radius $r = r_{CMB}$ (corresponding to the core-mantle boundary) is computed from the equation

$$k_c (dT/dr)_{ad,CMB} = k_c \alpha_c T_{CMB} g_{CMB} / c_{pc}, \qquad (8)$$

where the index "c" denotes the core parameters, and the index "CMB" the parameters taken at the core-mantle boundary. Note that if only a part of the core is convecting, the respective parameters are not taken at the CMB, but at the upper boundary of the involved zone. For the parameters see Table 1 and Table 2. We have tested the numerical code against existing codes in the literature. Details about these tests, the time discretization and the chosen resolution can be found in the Supplementary Material.

**3. Results**

In first sample runs with purely conductive heat transport (thus neglecting differentiation, heat transport by porous flow and convection in a magma ocean) we obtain a crude upper bound of the formation time $t_0$=2.9 Ma relative to the CAIs for the existence of a core - the solidus temperature of the silicate phase is exceeded in the centre of Vesta for formation times of up to 2.4 Ma and that of the metal phase for $t_0 \leq 2.9$ Ma relative to the CAIs (see Fig. 1). This upper boundary value can be refined if one considers differentiation.

Due to the effective radiogenic heating and the small surface-to-volume ratio the temperature in the mantle can reach several thousand degrees. Mantle temperatures exceeding the liquidus temperature of 1850 K of silicates are obtained for formation times of $t_0 \leq 2$ Ma after the CAIs but are inconsistent with the conclusions of Mason (1962), Stolper (1975,1977), Jones et al. (1984) and Jones et al. (1996), who suggest that the eucrites were generated by no more than 25% partial melting of the source region (noncumulative eucrite Sioux County, see Jones et al. 1996) and that 25% partial melting of the primordial mantle generated the crust. If one investigates the thermal evolution of Vesta without any additional cooling mechanisms except conduction, the only way to satisfy the physical constraints (e.g. 25% of silicate melting) is by assuming that Vesta formed at least 2 Ma after the CAIs. In this manner the thermal evolution of Vesta was investigated e. g. in Ghosh and McSween (1998). However, higher melt fractions could theoretically have been achieved in the source region after the extraction of melts which formed the eucrites.

Considering in a next step differentiation, partitioning of radioactive heat sources, heat transport by porous flow and convection, the thermo-chemical evolution differs significantly. Because $^{26}$Al is an incompatible element and is saturated in the liquid silicate phase during the melting of the silicates, the migration of the silicate melt has notable effects on the thermal evolution if this process is considered, because the concentration of $^{26}$Al becomes non-constant during and after the differentiation at different radii. Values of down to 0.003 have been suggested for the partitioning coefficient $P$, which controls the concentrations of $^{26}$Al in the solid and in the liquid silicate phase (Kennedy et al. 1993, Pack and Palme 2003). In the present study, we adopt the value $P=0.02$ to avoid a possible bias of the results.

The following scenario applies to a formation time $t_0<1.0$ Ma after the CAIs (see Fig. 2). In the early stages of the differentiation, when the first silicate melt is produced, $^{26}$Al-rich melt migrates towards the surface (at temperatures slightly above the silicate solidus of 1425 K) causing saturation of $^{26}$Al in a shallow sub-surface layer, whereas the solid matrix, which is depleted in $^{26}$Al, compacts downwards. In a shallow layer temperature increases rapidly due to the strongly enhanced production of radiogenic heat (see Fig. 3 and 4). Directly below a rigid undifferentiated shell the melt fraction rises up to the RCMF and convection starts in an approximately 1 km thick magma ocean. Close vicinity of the surface and effective cooling prevent the magma ocean from including the surface.

Directly above this superheated sub-surface layer and below the undifferentiated shell a region forms where silicate melt fractions lie below 50% and porous flow takes place. In this extremely thin boundary layer $^{26}$Al saturated melt percolates to the surface. Convection and efficient cooling through the surface causes the potential temperature in the magma ocean to vary around 1637 K (at this temperature 50% of the material in the silicate magma ocean is molten and thus the RCMF is crossed) resulting in a fluctuation between convective and advective regimes. In the advective regime the $^{26}$Al particles contained in the magma ocean migrate by porous flow into the boundary layer. This way, even more $^{26}$Al rises to the surface from the boundary layer together with the erupting silicate melt. Still, due to the differentiation which takes place further below, additional $^{26}$Al is transported into the magma ocean, which is subsequently heated above the critical temperature and switches again from the advective to the convective regime. Not all radioactive material leaves the mantle, nor does all of it reside at the surface. As soon as some portion of melt extrudes, it covers a part of the $^{26}$Al-enriched crustal material. We assume that this material transports $^{26}$Al into the boundary layer. This way material circulation between the surface and the boundary layer is established such that the critical temperature of 1637 K and thus a sub-surface magma ocean can be maintained right below the thin crust for as long as $O(10^4)$-$O(10^5)$ a. This mechanism leads to a substantial reduction of heating throughout the body as the major part of the radioactive material (except $^{60}$Fe) is involved in the material circulation between the crust and the upper mantle (here the crust is a solid basaltic carapace, the mantle is the layer with less than 1 vol% metal, the core is the central area with less than 1 vol% silicates). The sub-surface is the first completely differentiated part of the planetesimal. Simultaneously the differentiation of the deeper layers proceeds. Thereby the upper mantle is first cleared of the metal, then the rest of the mantle and the metal core form simultaneously.

The last stage of the silicate-metal differentiation is the formation of a distinct core-mantle boundary. After less than 0.3 Ma from the start of the melt migration, core formation is completed (e. g. for $t_0=0$ Vesta differentiates between 0.17 Ma and 0.33 Ma, for $t_0=0.5$ Ma between 0.78 Ma and 1 Ma, and for $t_0=1$ Ma between 1.5 Ma and 1.72 Ma). As a consequence of the depletion in $^{26}$Al of the mantle, the heat budget in almost entire body is reduced to the heating from the core by $^{60}$Fe decay only and the heating of the mantle from above by the magma ocean. Whereas the melt fraction in the magma ocean is fixed at about 50%, the melt fractions in the deeper regions remain lower (<40%, see Fig. 2 and 4). The effects of the sub-surface magma ocean on the interior temperature wear off after approximately 1 Ma. After this time the temperature profile throughout the mantle establishes a negative gradient from the CMB towards the surface (cooling of the core).

For models with $t_0>1.0$ Ma almost no melt extrusion at the surface is observed although for $1.0 \leq t_0 < 2.0$ a superheated layer in the sub-surface forms. However, note that we do not consider crust formation via dykes. For $t_0>2.0$ Ma the effects of partitioning of $^{26}$Al are negligible.

Regardless of the formation time, the melt fractions in the silicate mantle of Vesta remain below 40% (except in the shallow magma ocean), either due to the migration of the $^{26}$Al enriched melt towards the surface, or as a result of the decline in the radiogenic heating with time. No deep (whole-mantle) magma ocean is observed. Due to the density contrast between the silicate melt and the solid silicate matrix, some melt percolation takes place during the cooling of the depleted mantle. This stops when the temperature decreases below the silicate solidus, e. g., at maximally 150 Ma after the CAIs above the CMB. Silicate melt is present at a depth of 50 km until 28 Ma and at the depth of 100 km until 97 Ma after the CAIs if Vesta forms contemporaneously with the CAIs (see Fig. 2).

In the core, as thermal convection takes place only if the heat flow at the CMB surpasses the adiabatic heat flow at the CMB, considerably higher temperatures than the critical temperature for the metal phase of 1456 K resulting in 50% partial melt are attained. Up to 75% of the core material manage to melt before convection starts. The temperature profile in the convecting core is regulated by the temperature of the lower mantle directly above the CMB (due to the difference between the potential temperature and the temperature at the upper boundary, which enters the effective thermal conductivity, see Eqn. (5)-(7)). The typical values of the Rayleigh number here are $Ra=O(10^{19})$ and the effective thermal conductivity varies around $O(10^6)$ W m$^{-1}$ K$^{-1}$. The cooling through the CMB and the decline in the radiogenic heating puts the outer regions of the core into the advective regime below the RCMF of the metal phase. Subsequently the convecting inner part of the core becomes smaller through freezing of the outer regions until the convection ceases after ≈100 Ma after the formation of the body (Note, however, that we do not consider crystal segregation in a Fe-Ni-FeS core – the implications of this process on the results are discussed below).

Figure 6 shows the core radius (solid black line), the mantle thickness (dashed line) and the thickness of the undifferentiated outer layer (dotted line) as functions of the formation time. For $t_0 \leq 1.5$ Ma the core radius is approximately 111 km and is smaller for $1.5 < t_0 \leq 2.6$ Ma showing a sharp gradient between 2.3 Ma (97 km) and 2.6 Ma (≈1 km). The duration of the differentiation of rather small cores for the latter formation interval amounts to up to 170 Ma (as long as there is any melt). Formation prior to 1.5 Ma results in a complete or almost complete differentiation. At $t_0=2.7$-$2.8$ Ma no distinct core forms, although some increase in the metal concentration towards the centre can be observed as some melt migration below the radius of 216-188 km, respectively, took place. No melt migration occurred for $t_0=2.9$ Ma. As for $t_0>2.3$ Ma the solidus temperature of silicates was not exceeded, differentiation for the bodies with later formation time is due to iron melt migration only (and is much slower). For $t_0 \leq 2.3$ Ma a silicate mantle forms and for $t_0 \geq 1.5$ Ma a thin primordial (porous) layer is retained at the surface, thickness of the latter lying between 0.3 km at $t_0=1.5$ Ma and 2.8 km at $t_0=3$ Ma.

The present findings of a shallow magma ocean strongly depend on the migration velocity and here the most critical parameter is the silicate melt viscosity – the higher the viscosity the lower is the migration velocity and the more likely the formation of a global magma ocean and vice versa. In the present model, we have used a silicate melt viscosity of 1 Pas. The viscosity of molten silicates varies from 0.001 to 1000 Pa s (Moskovitz and Gaidos 2011), however, basaltic melts derived from chondritic precursors have silica contents generally ≲ 50% (Mittlefehldt et al. 1998) and viscosities of ≈1-100 Pa s (Giordano et al. 2008). Further tests with melt viscosities of 10 and 100 Pas show that an increase of the silicate melt viscosity results in lower melt migration velocities, thicker sub-surface magma ocean (e.g. ≈10 km for 10 Pa s in comparison to 1 km for 1 Pa s) and slower silicate separation. The cooling of the magma ocean below the RCMF is prolonged to 1 Ma and the layer is entirely crystallized after 3 Ma. Higher temperatures are obtained in the mantle and in the core, because of the somewhat prolonged radiogenic heating by the slow migrating $^{26}$Al. In the extreme case of 100 Pa s, the magma ocean even extends to the depth of ≈100 km. The rest of the mantle reaches melt fractions close to the RCMF and the iron-rich core, the silicate mantle and the crust from almost simultaneously. Apart from the thickness of the sub-surface magma ocean and the melt fractions (temperatures) in the body below, the most important effects described in the general scenario (fast silicate-metal differentiation, extrusion of $^{26}$Al, melt fractions in the magma ocean remain close to the threshold of 50%, etc.) are still valid.

The amount of heating by $^{60}$Fe is connected to the initial ratio $[^{60}Fe/^{56}Fe]_0=1.6 \times 10^{-6}$ adopted in our calculations. To test the influence of this parameter on the results we performed test runs assuming values of $1.0 \times 10^{-7}$, $1.0 \times 10^{-8}$ and $1.0 \times 10^{-9}$. The effect on the occurrence and evolution of the magma ocean is negligible, because the abundance of the major heat source $^{26}$Al remains constant. Furthermore, the melt fractions in the core and in the lower mantle decrease by less than 10% with the falling initial abundance of $^{60}$Fe. This decreases the duration of the thermal convection in the core slightly by less than 10 Ma.

In the following, we have also examined the influence of the partitioning of $^{26}$Al on the thermo-chemical evolution of Vesta by considering $P=1$ (no partitioning) to better compare the results with models that neglect partitioning. The duration of the differentiation is only slightly shorter than for $P=0.02$, but the order of crust-mantle and core-mantle differentiation differs. As in the case of $P=0.02$, first silicate and iron melt migration is observed at approximately 50 km depth and not in the centre - due to increasing gravity with radius and almost equal melt fraction as in the centre (owed to a flat temperature profile). Simultaneously, the silicate mantle grows down- and upwards from the region where the differentiation started, and the metal core grows upwards from the centre. In the last stage of the iron-silicate differentiation, a clear CMB is established. After core formation, a magma ocean starts to form somewhere in the mid-mantle. The melt from the boundary layer above the magma ocean is not able to percolate through the chondritic carapace and to form a basaltic crust. Even in the hottest scenario with $t_0$=0 Ma an undifferentiated (although compacted) primordial crust of ≈1 km thickness remains.

## 4. Discussion and Conclusions

Our simulations imply that no whole-mantle silicate magma ocean formed in the mantle of Vesta – percolation of silicate melt and the depletion of $^{26}$Al in the mantle is faster than the heating of the mantle by the radioactive elements, as also suggested by Moskovitz and Gaidos (2011) for smaller bodies. The entire silicate material did not exceed the critical temperature of approximately 1637 K (and the corresponding threshold melt fraction of 50%) and for the most part the melt fractions in the mantle remained below 40%. Even in the thin sub-surface magma ocean this critical temperature is exceeded by only few degrees. The critical temperature varies in general with the assumptions about the melting behavior of the silicate material (we assumed linear melting between the solidus and the liquidus of the silicates). Furthermore, the adopted value of the RCMF of 0.5 can be debated, as liquid-state convection may occur already at lower melt fractions (even values of down to 20% have been suggested, see e. g. Arzi 1978, Taylor 1992, Lejeune and Richet 1995, Scott and Kohlstedt 2006). Assuming a smaller threshold value of the RCMF for the onset of convection would result in the fixation of the temperature in the convecting zone at the temperature corresponding to the associated melt fraction. Thus, the source regions of the melt which formed the HED meteorites most likely never exceeded 20 to 50% melting and the corresponding temperature.

**Interior evolution of Vesta**

Based on our findings we suggest the evolution of the interior of Vesta shown in Fig. 7 **a**. After the initial accretion phase (which is approximated here by instantaneous formation at the time $t_0$), the radiogenic heating induces compaction of the porous material and the associated bulk volume decrease. Having a flat temperature profile with a sharp gradient in a thin outer porous layer, melt is produced even in shallow regions. Upon the onset of the differentiation $^{26}$Al is enriched in the outer 2-15 km of the body and the high temperatures and high degree of melting close to the surface cause compaction of the chondritic crust, the differentiation of the shallow layers and the melt eruptions at the surface. For viscosities of silicate melt between 1 and 10 Pas a shallow magma ocean with a thickness of 1 to 10 km forms. The magma ocean cools rapidly within $O(10^4)$-$O(10^6)$ a and its thickness as well as the time scale of solidification increases with increasing silicate melt viscosity.

Rapid cooling of the magma ocean on the HED parent body within 2-3 Ma has been suggested as a consequence of $^{26}$Mg variations among diogenites and eucrite (Schiller et al. (2011). However, Schiller et al. (2011) concluded that such a rapid cooling can occur only in asteroids with radii smaller than 100 km implying that Vesta is not the parent body of the HEDs. In contrast, our results admit both the size of Vesta and high cooling rates.

While $^{26}$Al is transported to the surface, the differentiation proceeds downwards: a basaltic crust, a purely silicate upper mantle and an iron-rich core form, with the completion of the mantle and a clear CMB subsequently. The observed basaltic surface of Vesta (e.g., McSween et al., 2013) favours formation times $t_0<1$ Ma, as in those cases the silicate melt is able to penetrate the outermost layer via porous flow and thereby to cover the entire surface. This correlates with the lack of a large amount of relict undifferentiated material (e.g., De Sanctis et al. 2012) and the presence of basaltic crust but lack of distinct volcanic regions on Vesta (Jaumann et al. 2012). Volcanism caused by melt transport via dykes may violate these findings. Furthermore, the $^{26}$Mg composition of the most primitive diogenites requires onset

630  of the magma ocean crystallization within $0.6^{-0.4}_{+0.5}$ Ma after CAI's (Schiller et al., 2011)

631  consistent with the formation times of $t_0$<1 Ma.

632  The silicate-metal differentiation is rapid and in general completed within 0.3 Ma after

633  initiation of melting; nevertheless it succeeds early silicate differentiation. This sequence is

634  different to what is generally assumed for planetesimals from chronological records, i.e., core

635  formation precedes crust formation (e.g., Kleine et al., 2009). To reverse the differentiation

636  sequence either metal segregation needs to be accelerated or silicate segregation slowed

637  down. Metal segregation is enhanced for instance for a high concentration of FeS close to the

638  eutectic (Neumann et al., 2011) or a lower viscosity of iron melt. In the present study, we

639  have assumed a L-chondritic composition of Vesta with only 5 wt% of FeS and a melt

640  viscosity of $10^{-2}$ Pas. However, higher concentrations of FeS in Vesta have been also

641  suggested with values of ~10 wt%, corresponding to a Vesta-like body composed of 70 wt%

642  ordinary chondrite and 30 wt% carbonaceous chondrite (Boesenberg and Delaney, 1997).

643  Furthermore, experiments show a variation of the viscosity of iron-rich melt between $10^{-3}$ and

644  $10^{-2}$ Pas (Dobson et al., 2000; Assael et al., 2006) and we took the standard value of $10^{-2}$ Pas

645  used in the literature.

646  Although we cannot discriminate the composition of the various silicate liquid and solid

647  phases in our model, we speculate the first partial melt reaching the surface are non-

648  cumulative eucrites. The cumulative eucrites and the diogenites are suggested to have formed

649  as a consequence of the crystallising sub-surface magma ocean. The latter has an olivine-poor

650  composition as it contains mainly partial melt that is transported upward from the mantle

651  below. In addition, at the same time the original solid matrix in the magma ocean region is

652  pushed down and replaced by crystallising partial melt from the lower mantle. Thus, in our

653  scenario the non-cumulative eucrites form prior to the cumulative eucrites and diogenites.

654  This order of formation is supported by the internal $^{26}$Al-$^{26}$Mg isochrones obtained by Hublet et al. (2013), who infer younger ages for the diogenites compared to those of the eucrites, and younger ages for the cumulative eucrites compared to the non-cumulative ones.

It should be noted, though, that silicate melting and melt migration is ongoing for up to 150 Ma and thus much longer than silicate-metal separation. However, the upper melting front moves continuously to greater depth. Partial melting generated later than the early eucritic crust is not able to rise to the surface and will remain in the lower crust or mantle.

High values of the silicate melt viscosity between 10 to 100 Pas suggest that the subsurface magma ocean can even reach a thickness up to 100 km. A sub-surface magma ocean with a thickness of a few tens of kilometres would in fact be more consistent with the assumption of Vesta's crust having an average thickness of 35-85 km with an upper eucritic and a lower orthopyroxene-rich layer (McSween et al., 2013).

Our results suggest that no global magma ocean down to the core mantle boundary forms because the silicate melt including $^{26}$Al is transported toward the surface. This result is in particular valid for a RCMF of 0.5. For smaller values of the critical melt fraction - that determines the change in the rheological behaviour and defines the existence of a magma ocean from the physical perspective (low viscosity) - the critical temperature is reached faster. However, in our approach we have neglected the composition of the various silicate liquid and solid phases and thereby also the variation in the melting temperature with composition. Depleting the mantle in crustal components due to magma migration, results in higher solidus temperatures of the residual material and thus reduced subsequent partial melting in the lower mantle. In contrast, in the shallow magma ocean, which accumulates partial melt of the lower mantle that has a lower liquidus temperature than the primordial material, a higher degree of melting is possible (than calculated in our approach). Thus, a more detailed study that also considers the individual silicate phases may strengthen our finding of a shallow sub-surface and no global magma ocean on Vesta.

The iron-rich core starts to convect by thermal buoyancy already during its formation. Convection and cooling of the core through the depleted mantle is then maintained for up to $O(10^8)$ a. It should be noted, though, that the core was - like the mantle - not entirely melted and the melt fraction stayed below 75%. Thus, an early dynamo is not by thermal convection only, i.e. a purely early thermal dynamo in not likely in Vesta. Chemical convection due to freezing processes in the core may play a substantial role for the magnetic field generation in planetesimals (Nimmo, 2009). The freezing process and chemical convection has been simplified in the present study and needs further studies.

The less realistic evolution path from Fig. 7 b results in the formation of a convecting whole-mantle magma ocean. Here, the distribution of $^{26}$Al is homogeneous throughout the silicate phase (both solid and liquid). Note that by contrast to previous thermo-chemical evolution models considering a magma ocean scenario (e.g. Gupta and Sahijpal, 2010), the differentiation does not start in the centre, but in the upper mantle. Furthermore, the silicate melt from the mantle is not able to penetrate the chondritic crust via porous flow, thus likely no basalts form at the surface and the differentiation is not complete. However, ongoing impacts, presence of volatiles, and excess pressure in the interior due to melting could act in favour of melt ascent via cracks to produce a basaltic crust (Wilson and Keil, 2012). The chondritic crust could be also carved off by impacts exposing the differentiated silicate layers although one would expect relicts of undifferentiated material that have not been observed (e.g., De Sanctis et al. 2012).

**Comparison to previous thermo-chemical evolution models**

Our results suggest that most previous thermo-chemical evolution models for Vesta tend to overestimate the temperature increase in the interior and therefore the amount of partial melting (Righter and Drake 1997, Ghosh and McSween 1998, Drake 2001, Gupta and Sahijpal 2010). Those models either neglect efficient heat transport by the partitioning of $^{26}$Al into the silicate melt or convection in a magma ocean.

The convective cooling method used by Hevey and Sanders (2006) for small planetesimals < 100 km in radius (increase of $k$ by three orders of magnitude for melt fractions ≥50%) has been also adopted by Gupta and Sahijpal (2010) for Vesta (a steady increase of the thermal diffusivity between the melt fractions of 50% and 100% by three orders of magnitude). Thus, a maximal value of ≈3300 W m$^{-1}$ K$^{-1}$ has been used for effective thermal conductivity by Gupta and Sahijpal (2010). The effective thermal conductivity, however, varies with the cube root of gravity times the depth of the convecting zone (Eqn. (5) - (7)). Hence, the larger the radius and / or the convecting zone, the larger is $k_{eff}$. In our simulations values of $O(10^6)$ W m$^{-1}$ K$^{-1}$ can be obtained for $\Delta T = O(1)$ K (the difference between the potential temperature and the temperature at the upper boundary of the convecting zone. This explains the much higher temperatures of up to 2000 K obtained by Gupta and Sahijpal (2010). Our model in which the partitioning of $^{26}$Al is neglected as in Gupta and Sahijpal (2010) show a global magma ocean but still the temperatures do not exceed significantly the threshold melt fraction of 50% due to efficient cooling.

Our results further imply that previous thermal evolution models for planetesimals to explain the magnetic field generation in their iron-rich cores such as presented in e. g. Elkins-Tanton et al. (2011), Elkins-Tanton (2012) and Sterenborg and Crowley (2013) have to be considered with care although they consider efficient cooling by convection. These models start with a differentiated structure and further neglect volcanic heat transport and partitioning of $^{26}$Al. Thereby, thermal convection in the core and magnetic field generation is initiated due the efficient cooling of the magma ocean after an initial phase of heating by the radioactive heat sources. It can be shown in our models that when considering the entire differentiation process, i.e. silicate-metal and silicate-silicate differentiation, that the temperature profile after core formation shows a temperature profile increasing toward the surface and inhibiting convection. A whole-mantle magma ocean does not form and, as a consequence, the lower mantle cools by conduction only. Nevertheless, the core heat flow is higher than the heat flow along the core adiabat suggesting thermal convection in the core for about 100 Ma. This finding is in contrast to the assumption of Sterenborg and Crowley (2013), who argue that thermal convection in the core is inhibited when the mantle cools by conduction using a simple model of conductive cooling. In should be noted though that the strength of thermal core convection in our model is likely not sufficiently strong to generate a dynamo, i.e. is the magnetic Reynolds number is lower than about O(10-100) (e.g., Christensen et al., 1999). However, we do not explicitly differentiate between the iron-nickel system and lighter elements in the core (such as those contained in the Fe-FeS system). This means that our model does not allow freezing and growth of a solid inner core and the establishment of a liquid outer core due to the exclusion of light elements from the crystal structure, as well as the latent heat release at the inner core boundary. Thus, chemically driven convection (Stevenson 2003; Nimmo, 2009) and its contribution to the magnetic field are neglected.

**Solid-state convection**

As mentioned further above, the solid-state plastic deformation (other than compaction due to sintering, see e.g. Tkalcec et al. 2013) for melt fractions below 50% has been neglected in the present study. First, the conditions for solid-state convection are fulfilled several million years after the solidification of the shallow magma ocean. Before that time, the temperatures in the upper mantle (magma ocean region) are higher than in the lower mantle and thermal convection is suppressed. Second, the maximal Rayleigh number that we obtain when the conditions are favourable for convection is estimated to be ≈5500, which exceeds the critical Rayleigh number of 1000 (e.g. Schubert et al. 2001) required for convection only moderately. Because this type of convection takes place after the magma ocean has vanished, it has no effect on the main results presented in the paper. However, the latest time of occurrence of partial melt in the mantle and the time scale of the convection in the core could change slightly.

**Melt ascent through a porous crust and formation of dykes**

Our results suggest the formation of a basaltic crust by porous flow. It has been argued however, that melt ascent through a porous chondritic crust is unlikely due to the lack of the positive density contrast (e.g. Elkins-Tanton et al. 2011). In our simulations of Vesta the porous chondritic material compacts readily due to hot pressing if exposed to the temperatures far below the silicate solidus of ≈1425 K. Consequently, the density contrast between the melt and the matrix is still maintained. This result is supported by the numerical simulations of Henke et al. (2012, 2013). When the silicate melt reaches the porous layer, this layer has already been exposed to temperatures high enough to compact (and compacted in the course of a self-consistent calculation). Eventually, an extremely thin surface shell (~100 m thick) remains porous which does not represent a barrier for buoyant rising melt (see Fig. 6 and discussion on grid resolution in the Supplementary Material). On the other hand, compacted chondritic material at the surface could have been fractured due to impacts. However, this should not hinder melt ascension, but would rather favour the extraction of melt via dykes. Also, large pressures can be generated in the sub-surface for small degrees of melting (e. g. a pressure of 0.1 GPa for a melt fraction of 0.2, see Wilson and Keil, 2012), and eruptions would occur when the excess pressure due to the accumulation of magma in the sub-regolith reservoirs exceeds the tensile stress of the overlying rock (Wilson and Keil, 2012). The presence of small amounts of volatiles reduces the magma density and increases even the excess pressure. Note, though, that to keep the fractures open and to allow efficient cooling by the silicate melt transport directly from the deep mantle to the surface, Vesta may be a boundary case due to its small size. Instead, magma would first accumulate in sills or magma layers at the base of the upper thermal boundary layer before eruption (Wilson and Keil, 2012).

Efficient melt migration via dykes – if possible - suggests even faster cooling of the interior than via porous flow and even strengthens the result presented here. Consequently, faster cooling of the mantle and a shorter time span of the convection in the core are expected.

979

980

981

982

983

984

985

986

987

988

989

990

991

992

**Figure captions**

Fig. 1. Occurrence of iron and / or silicate melt for different formation times of Vesta assuming heat transport by conduction only and no differentiation.

Fig. 2. Temporal evolution of the radial distribution of the interior temperature in K (left column) and melt fraction (right column) for different formation times relative to the CAIs. The panels correspond from top to bottom to $t_0$=0, 1, 2, and 2.3 Ma.

Fig. 3. Temperature (in K) ant two close-ups in the sub-surface magma ocean of a body which formed at $t_0$=0.5 Ma after the CAIs.

Fig. 4. Melt fraction and two close-ups in the sub-surface magma ocean of a body which formed at $t_0$=0.5 Ma after the CAIs.

Fig. 5. Close-ups of the sub-surface magma ocean or the super-heated sub-surface region for different formation times: $t_0$=0 Ma (upper panels), 1 Ma (middle panels), 2 Ma (lower panels). Left column: Temperature is in K. Right column: Melt fraction.

Fig. 6. Left panel: Density versus depth for different formation times after the cooling below the solidus temperature of the metal phase. Right panel: Core radius, mantle thickness and the thickness of the undifferentiated layer for different formation times. Note that for $t_0 \geq 2.3$ Ma no clear mantle forms.

Fig. 7. **a** Evolution path of the interior assuming partitioning of $^{26}$Al. **b** Evolution path of the interior neglecting partitioning of $^{26}$Al.

## Tables

**Table 1.** Composition, density and thermal conductivity.

| Phase | Mineral | Mass fraction [%] | Volume fraction [%] | $\rho_{solid}$ [kg m$^{-3}$] | $\rho_{liquid}$ [kg m$^{-3}$] | k [W m$^{-1}$ K$^{-1}$] |
|---|---|---|---|---|---|---|
| **Metal** | - | 14 | 7.79 | 6471.12 | 5490.88 | 10.2 |
| | Fe | 8 | - | 7874 | 6980.00 | 21 |
| | Ni | 1 | - | 8908 | 7810.00 | 21 |
| | FeS | 5 | - | 4830 | 3920.00 | 4.6 |
| **Silicates** | - | 86 | 92.21 | 3357.55 | 2900 | 3.8 |
| | Olivine | 49 | - | 3510.00 | - | 4.3 |
| | orthopyroxene | 23 | - | 3380.00 | - | 3.9 |
| | clinopyroxene | 6 | - | 3320.00 | - | 4.6 |
| | Plagioclase | 8 | - | 2630.00 | - | 1.9 |
| **Total** | **L-chondrite** | **100** | **100** | **3600.05** | **3105.12** | **4.2** |

This data is for the most part from Yomogida and Matsui 1983 for L-chondrites.

**Table 2.** Parameter values used for the models.

| Quantity | Symbol | Unit | Value | Reference |
|---|---|---|---|---|
| Initial porosity | $\phi_0$ | - | 0.4 | 1 |
| Reference radius | $D$ | Km | 262 | 2 |
| Initial radius | $R_p$ | Km | 310 | 2 |
| Thermal expansivity of iron | $\alpha_{Fe}$ | K$^{-1}$ | 7.70x10$^{-5}$ | 3 |
| Thermal expansivity of silicates | $\alpha_{Si}$ | K$^{-1}$ | 2.00x10$^{-5}$ | 4 |
| Iron melt viscosity | $\eta_{l,Fe}$ | Pa s | 10$^{-2}$ | 1 |
| Silicate melt viscosity | $\eta_{l,Si}$ | Pa s | 1 | 1 |
| Metal solidus temperature | $T_{S,Fe}$ | K | 1213 | 1 |
| Metal liquidus temperature | $T_{L,Fe}$ | K | 1700 | 1 |
| Silicate solidus temperature | $T_{S,Si}$ | K | 1425 | 1 |
| Silicate liquidus temperature | $T_{L,Si}$ | K | 1850 | 1 |
| Radius of iron globules | $b$ | M | 0.01 | 2 |
| Partitioning coefficient | $P$ | - | 0.02 | 1 |
| Stefan-Boltzmann constant | $\sigma_{SB}$ | W m$^{-2}$ K$^{-4}$ | 5.67x10$^{-8}$ | - |
| Emissivity | $E$ | - | 0.8 | 5 |
| Avogadro number | $N_A$ | mol$^{-1}$ | 6.02x10$^{23}$ | - |
| Nebula temperature | $T_N$ | K | 290 | 1 |
| Decay energy per atom of $^{26}$Al | $E_{26Al}$ | J | 6.41·10$^{-13}$ | 1 |
| Decay energy per atom of $^{60}$Fe | $E_{60Fe}$ | J | 4.87·10$^{-13}$ | 1 |
| Initial abundance of $^{26}$Al | $f_{26Al}$ | kg$^{-1}$ | 2.76·10$^{23}$ | 2 |
| Initial abundance of $^{60}$Fe | $f_{60Fe}$ | kg$^{-1}$ | 1.12·10$^{24}$ | 2 |
| Mass fraction of bulk Al | $m_{Al}$ | % | 1.19 | 2 |
| Mass fraction of bulk Fe | $m_{Fe}$ | % | 11.18 | 2 |
| Initial ratio of $^{26}$Al | [$^{26}$Al /$^{27}$Al]$_0$ | - | 5·10$^{-5}$ | 1 |
| Initial ratio of $^{60}$Fe | [$^{60}$Fe /$^{56}$Fe]$_0$ | - | 1.6·10$^{-6}$ | 1 |
| Half-life of $^{26}$Al | $\tau_{1/2}(^{26}Al)$ | Ma | 0.717 | 1 |
| Half-life of $^{60}$Fe | $\tau_{1/2}(^{60}Fe)$ | Ma | 2.61 | 6 |

**References:** 1 Neumann et al. 2012. 2 Model section. 3 Rivoldini et al. 2011. 4 Sohl and Spohn 1997. 5 Ghosh and McSween 1998. 6 Rugel et al. (2009)

Figures

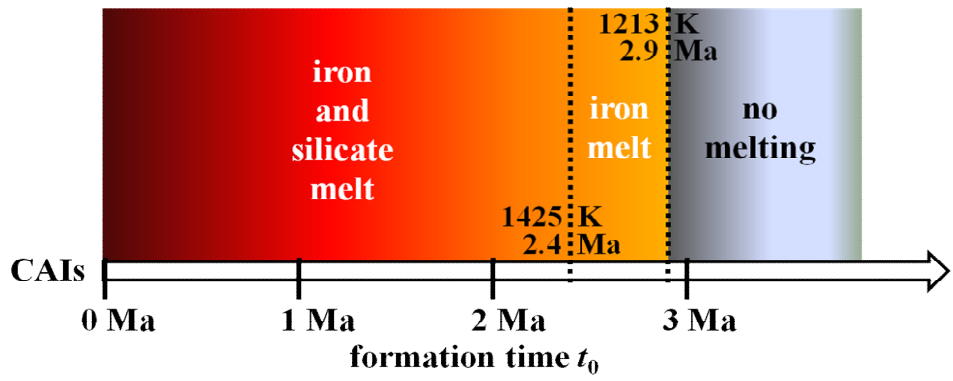

**Fig. 1.**

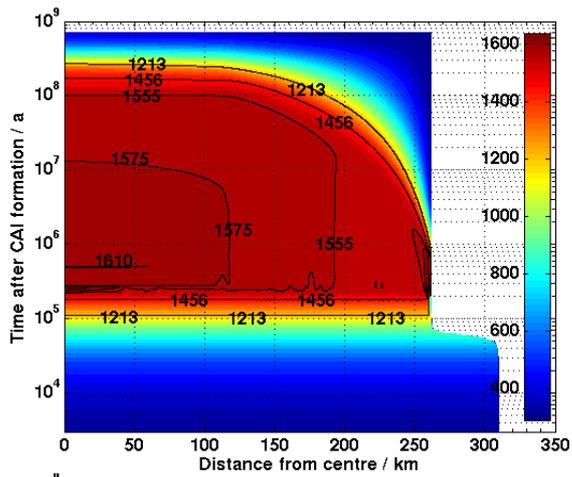 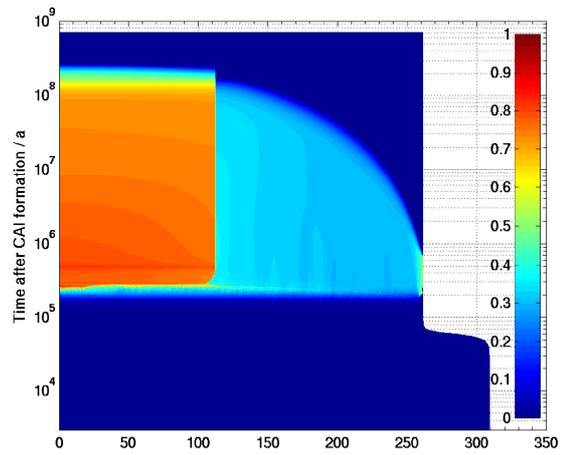
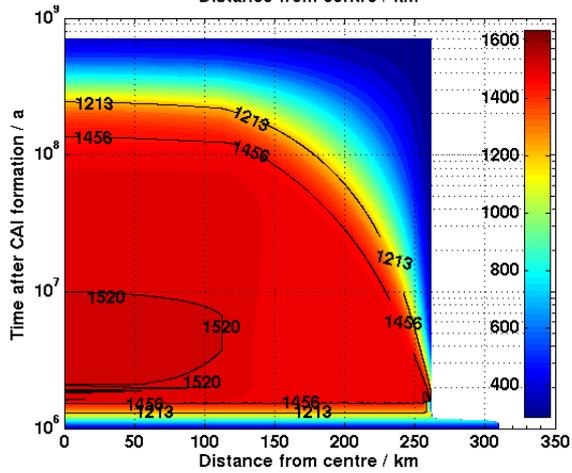 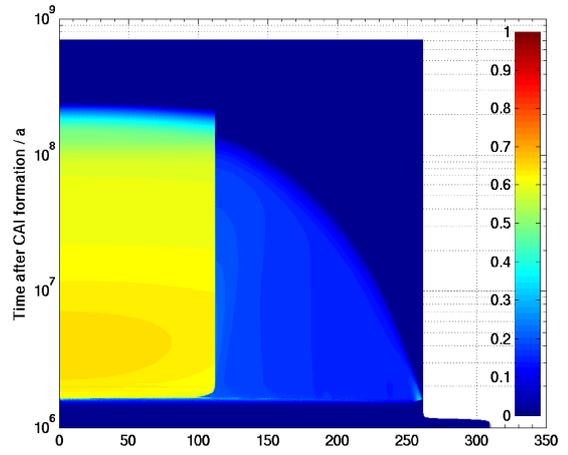
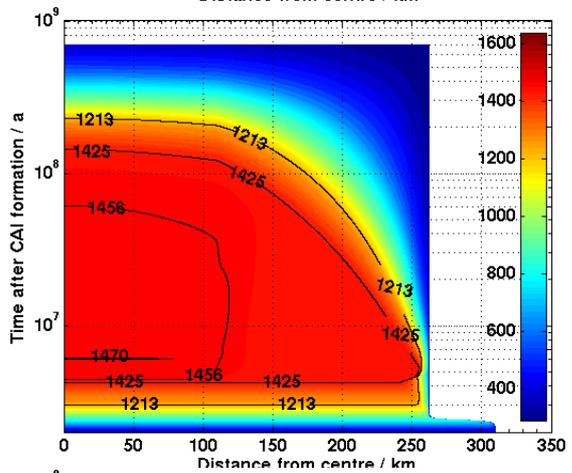 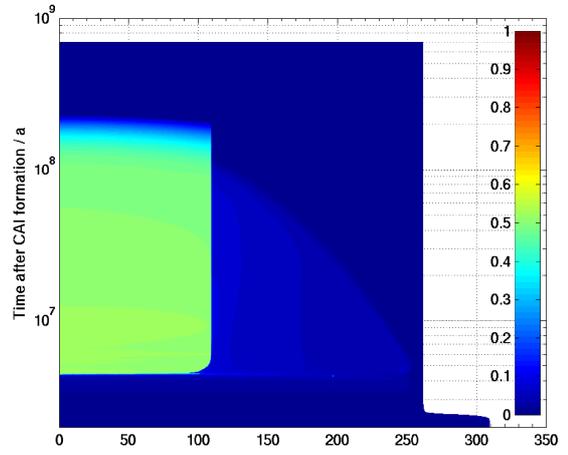
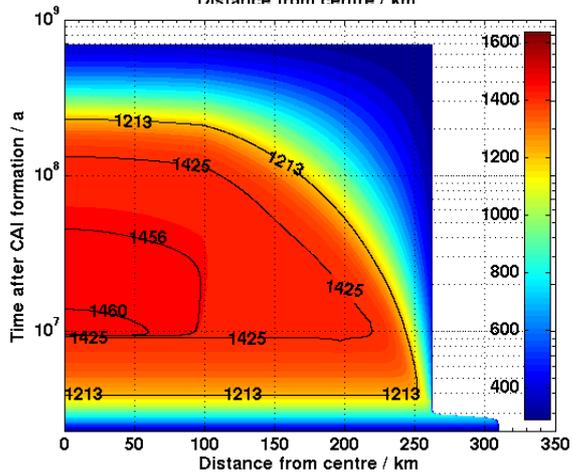 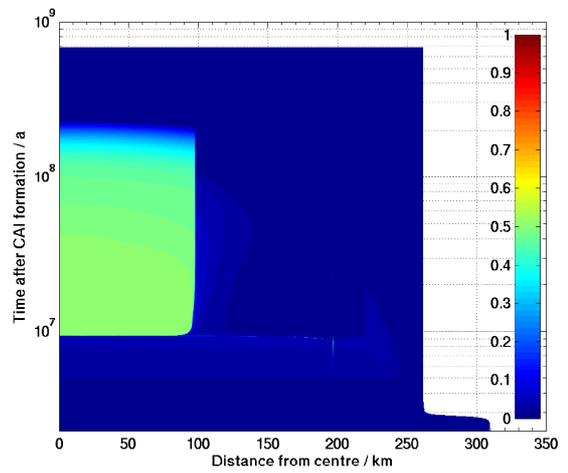

1035    **Fig. 2.**

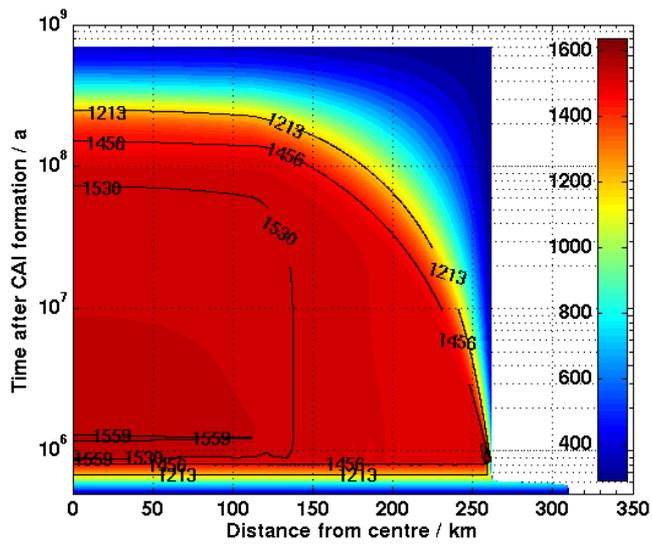
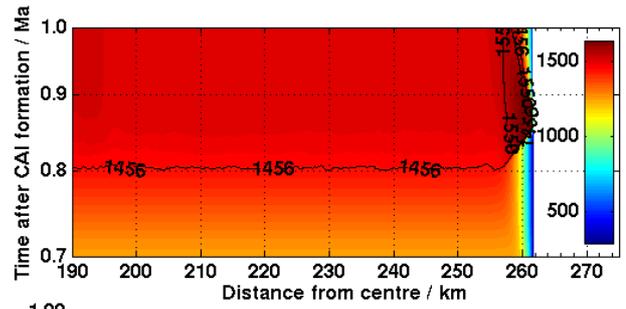
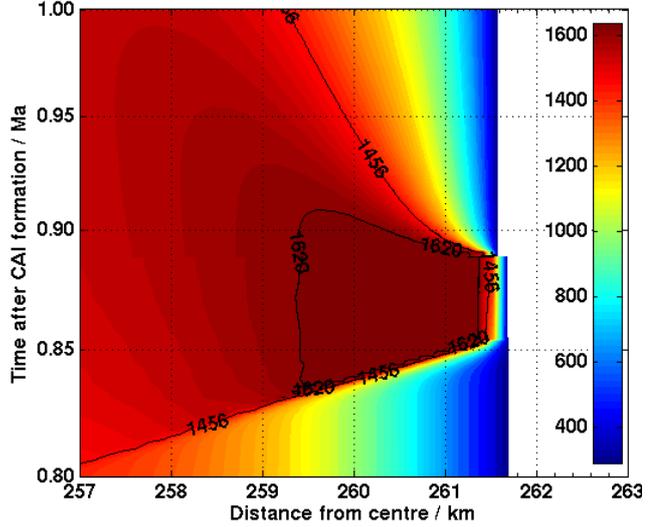

**Fig. 3.**

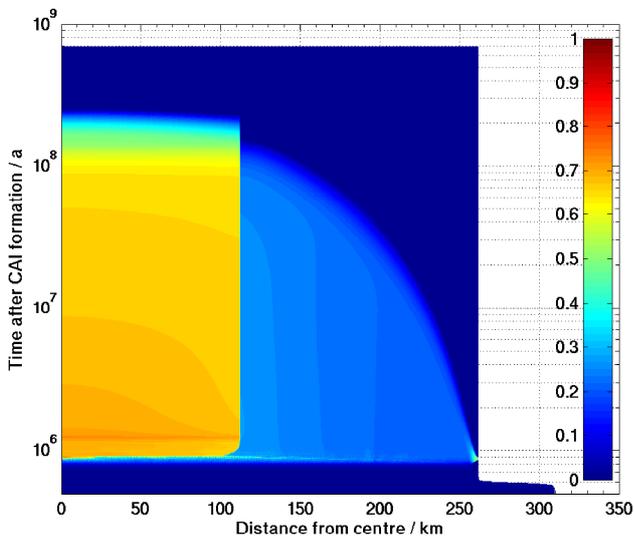
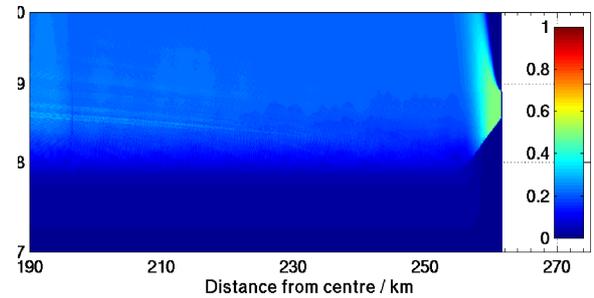
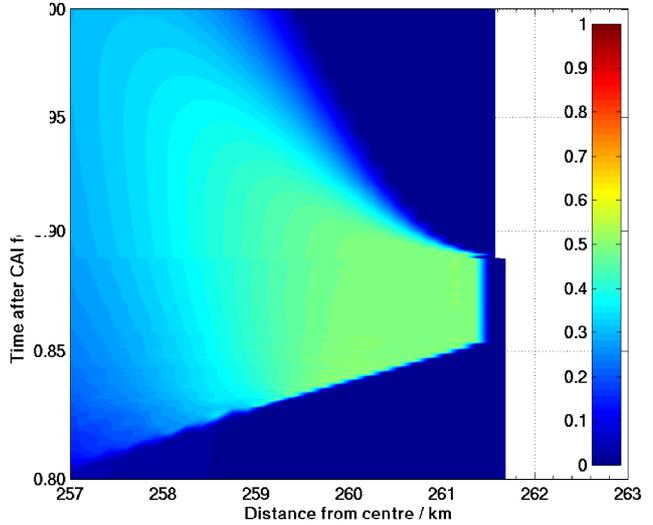

**Fig. 4.**

1044

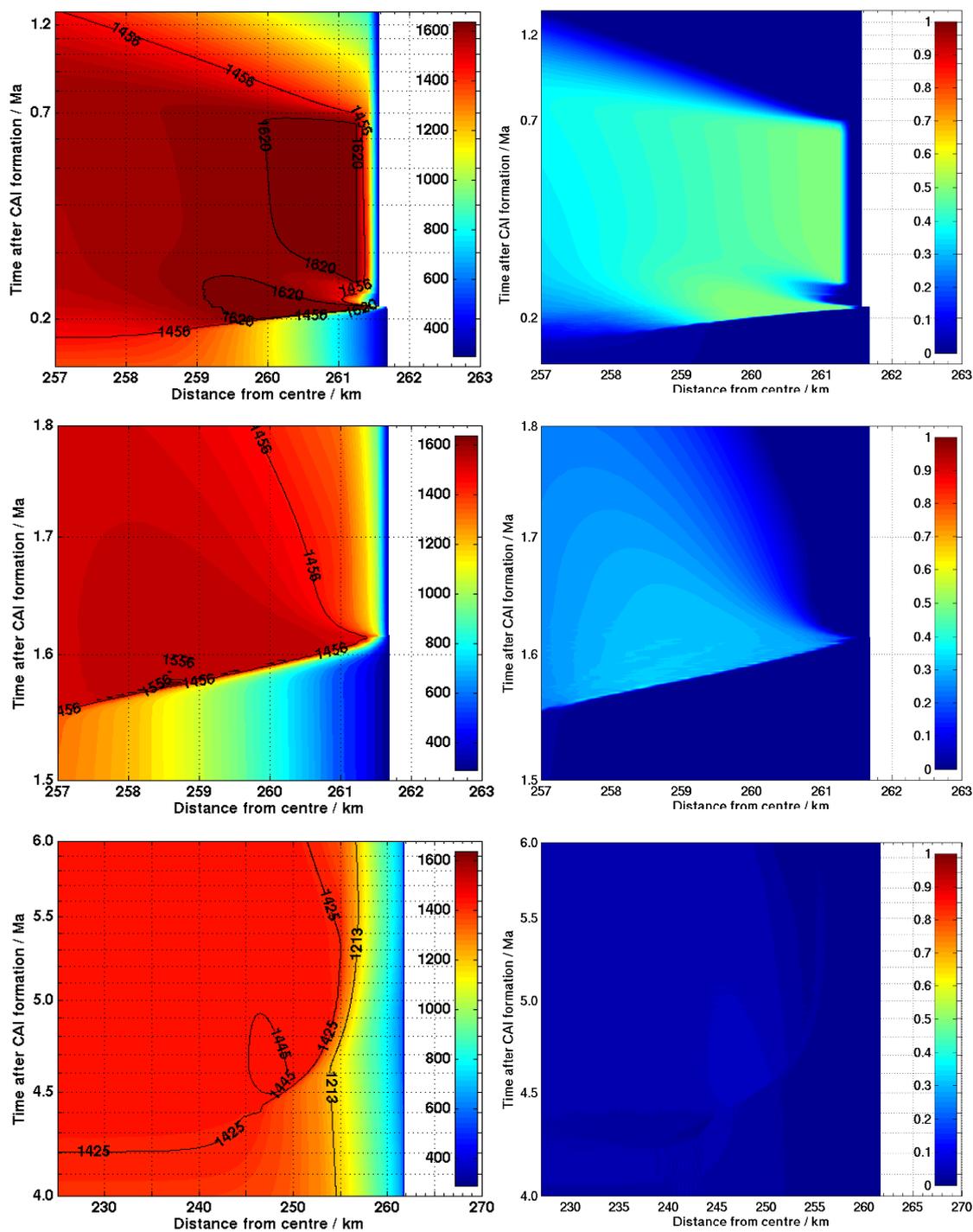

1045 **Fig. 5.**

1046

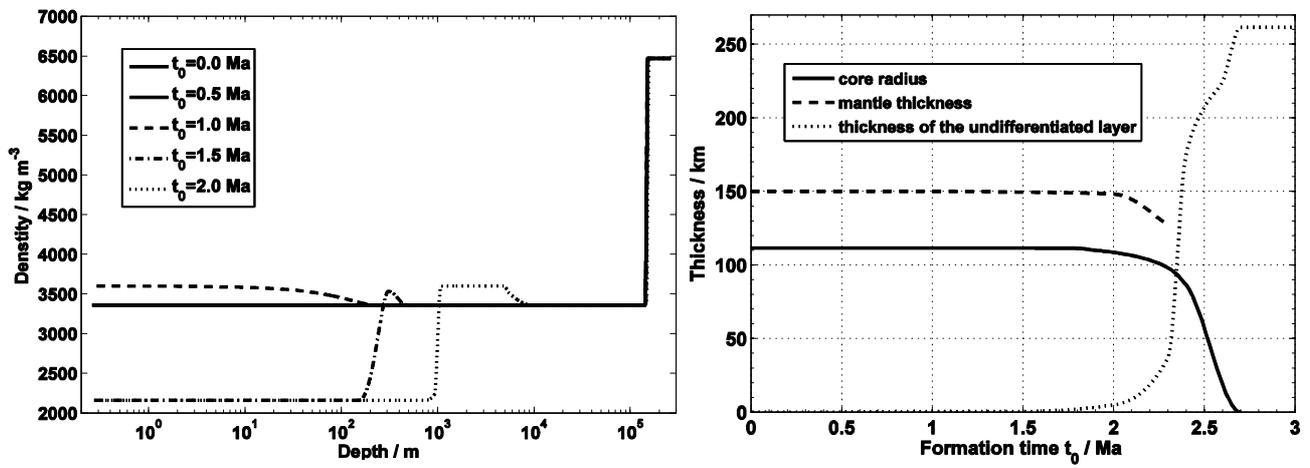

**Fig. 6.**

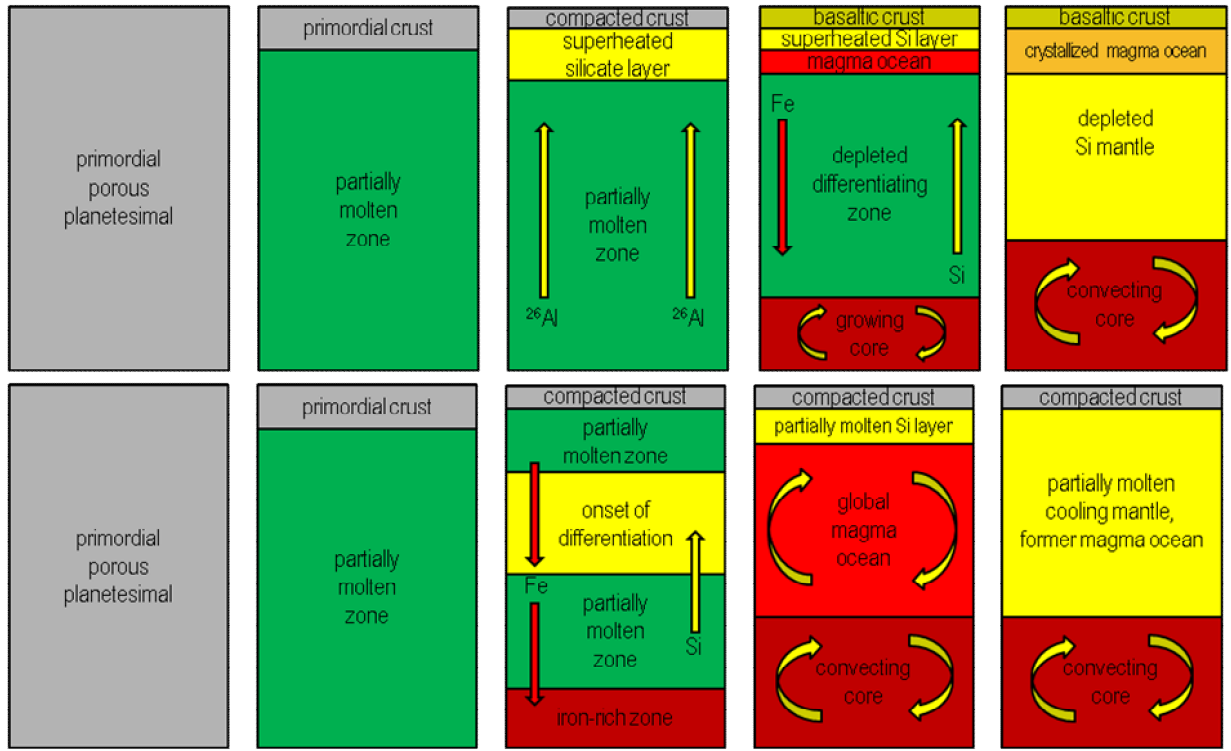

**Fig. 7.**

# Supplementary material

**Benchmark tests**

In the following we will show tests of our numerical approach against existing models in the literature. It should be noted, however, that there are not many codes which consider additional processes apart from conduction in planetesimals. Among those, only a few are available or are documented well enough for us to be able to recalculate some of the results. Furthermore, in most cases the processes like melting, consumption and release of latent heat, and differentiation are treated in a completely different manner. We compare heat conduction (Miyamoto et al. 1981, Šrámek et al. 2012), melting and latent heat consumption and release (Moskovitz 2011, Merk 2002), and differentiation (Šrámek et al. 2012).

**1 Miyamoto et al. (1981) (conduction only)**

Parameters for an H-chondrite adopted by Miyamoto (1981):

| | | | |
|---|---|---|---|
| Radius | $R$ | 85 | km |
| Formation time rel. to CAIs | $t_0$ | 0 | Ma |
| Thermal conductivity | $k$ | 1 | W m$^{-1}$ K$^{-1}$ |
| Density | $\rho$ | 3200 | kg m$^{-3}$ |
| Specific heat capacity | $c_p$ | 625 | J kg$^{-1}$ K$^{-1}$ |
| Initial / surface temperature | $T_0, T_N$ | 200 | K |
| Porosity | $\phi$ | 0 | - |
| Initial ratio $^{26}$Al/$^{27}$Al | $[^{26}$Al/$^{27}$Al$]_0$ | 5.0x10$^{-6}$ | - |
| Initial heat production | $Q_0$ | 11.67*$[^{26}$Al/$^{27}$Al$]_0$ | W m$^{-3}$ |
| Half-life of $^{26}$Al | $t_{1/2}$ | 0.72 | Ma |
| Decay energy per atom | $E_{Al}$ | 5.1x10$^{-13}$ | J |
| Aluminum mass fraction | $m_{Al}$ | 0.0101 | - |

Thereby, the initial heat production per 1 m$^3$ follows from the aluminium mass fraction of 0.0101 (see the reference Loveland 1969 in Miyamoto et al. 1981).

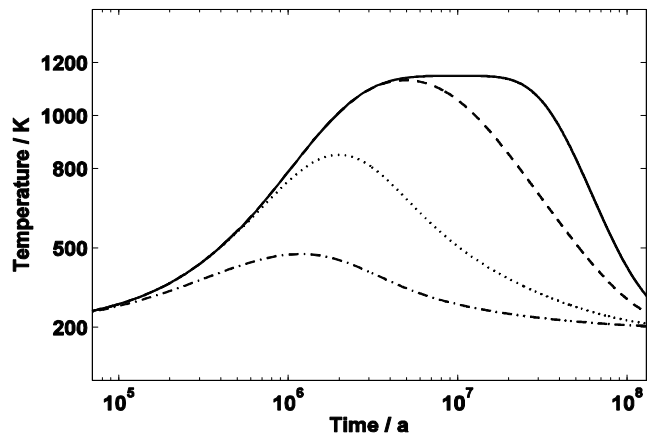

**Fig. S.1:** Temperature evolution at different depths. The lines correspond to the distances from centre of 0 (solid), 55 (dashed), 77 (dotted) and 82.5 (dot-dashed) km. The lines coincide with those of Figure 1 (a) from Miyamoto et al. (1981).

## 2 Šrámek et al. 2012 (conduction only)

Parameters

| | | | |
|---|---|---|---|
| Radius | $R$ | 500 | km |
| Formation time rel. to CAIs | $t_0$ | 0, 1, 2 | Ma |
| Fe volume fraction | $v_{Fe}$ | 0.18 | - |
| Si volume fraction | $v_{Si}$ | 0.82 | - |
| Fe thermal conductivity | $k_{Fe}$ | 50 | W m$^{-1}$ K$^{-1}$ |
| Si thermal conductivity | $k_{Si}$ | 3 | W m$^{-1}$ K$^{-1}$ |
| Fe density | $\rho_{Fe}$ | 7800 | kg m$^{-3}$ |
| Si density | $\rho_{Si}$ | 3200 | kg m$^{-3}$ |
| Fe specific heat capacity | $c_{p,Fe}$ | 450 | J kg$^{-1}$ K$^{-1}$ |
| Si specific heat capacity | $c_{p,Si}$ | 1200 | J kg$^{-1}$ K$^{-1}$ |
| Initial / surface temperature | $T_0, T_N$ | 300 | K |
| Porosity | $\phi$ | 0 | - |
| Initial ratio $^{26}$Al/$^{27}$Al | $[^{26}$Al/$^{27}$Al$]_0$ | 5.0x10$^{-5}$ | - |
| Initial heat production | $Q_0$ | 1.5x10$^{-7}$ | W kg$^{-1}$ |
| Half-life of $^{26}$Al | $t_{1/2}$ | 0.717 | Ma |
| Decay energy per atom | $E_{Al}$ | 5.1x10$^{-13}$ | J |
| Aluminum mass fraction | $m_{Al}$ | 0.00865 | - |

Thereby, the average values of the mixture are calculated as volume fraction weighted averages of the respective values of Fe and Si.

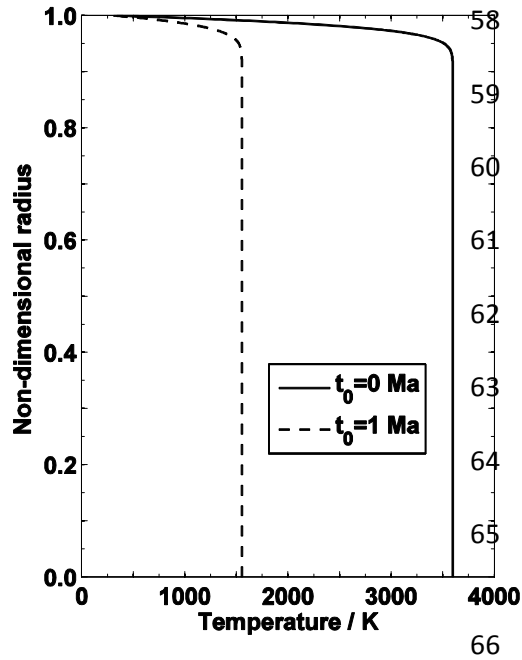

**Fig. S.2:** Results of the computations performed using our code with the parameters from above. The lines show temperature as function of normalised radius at 1 Ma after the formation. The formation times are $t_0 = 0$ Ma (solid line) and $t_0 = 1$ Ma (dashed line) relative to the formation of the CAIs. The temperature profiles are identical to those in Fig. 1 in Šrámek et al. 2012 (thin solid and thin dashed lines).

## 3 Moskovitz and Gaidos (2011) (melting, latent heat consumption and release, no differentiation)

Parameters

| | | | |
|---|---|---|---|
| Radius | $R$ | 50 | km |
| Formation time rel. to CAIs | $t_0$ | 1 | Ma |
| Thermal conductivity | $k$ | 2.1 | W m$^{-1}$ K$^{-1}$ |
| Density | $\rho$ | 3300 | kg m$^{-3}$ |
| Matrix specific heat capacity | $c_{p,s}$ | 800 | J kg$^{-1}$ K$^{-1}$ |
| Melt specific heat capacity | $c_{p,m}$ | 2000 | J kg$^{-1}$ K$^{-1}$ |
| Initial / surface temperature | $T_0, T_N$ | 180 | K |
| Porosity | $\phi$ | 0 | - |
| Initial abundance of $^{26}$Al | $f_{Al}$ | 2.62x10$^{23}$ | kg$^{-1}$ |
| Initial ratio $^{26}$Al/$^{27}$Al | $[^{26}Al/^{27}Al]_0$ | 5.0x10$^{-5}$ | - |
| Decay energy per atom | $E_{Al}$ | 6.4154x10$^{-13}$ | J |
| Half-life of $^{26}$Al | $t_{1/2}$ | 0.74 | Ma |
| Initial abundance of $^{60}$Fe | $f_{Fe}$ | 2.41x10$^{24}$ | kg$^{-1}$ |
| Initial ratio $^{60}$Fe /$^{56}$Fe | $[^{60}Fe/^{56}Fe]_0$ | 6.0x10$^{-7}$ | - |
| Decay energy per atom | $E_{Al}$ | 4.87x10$^{-13}$ | J |
| Half-life of $^{60}$Fe | $t_{1/2}$ | 2.62 | Ma |
| Solidus temperature | $T_S$ | 1373 | K |
| Liquidus temperature | $T_L$ | 2009 | K |
| Latent heat of Si | $L$ | 400 | kJ kg$^{-1}$ |

Above $T_L$ the effective specific heat capacity is computed as melt fraction weighted average of the heat capacities of the matrix and the melt. Thereby, the melt fraction $\chi$ is calculated from an empirical fit to the fertility function of peridotite (see reference McKenzie and Bickle 1988 in Moskovitz and Gaidos 2011):

$$\chi(T',T) = 0.3936 + 0.253 \cdot T' + 0.4256 \cdot T'^2 + 2.988 \cdot T'^3,$$

where

$$T'(T) = (T-1691)/636.$$

Here, $636 = T_L - T_S$, and $1691 = (T_S + T_L)/2$. Note that although $(T_S + T_L)/2 = 1418.1$ is stated in Moskovitz and Gaidos (2011), the correct value of 1691 follows from McKenzie and Bickle (1988) and, as far as our calculations show, the correct value seems to have been adopted in Moskovitz and Gaidos (2011).

We have used this fit $\chi(T')$ to produce melt in our model. The derivative $d\chi(T')/dT$ represents the melting rate. This was used to derive the Stefan number (instead of using the melting rate $1/(T_L - T_S)$ which follows from the linear melting which is assumed in our model). No latent heat consumption or release due to metal melting or solidification was assumed, as the model by Moskovitz and Gaidos (2011) considers a homogeneous body.

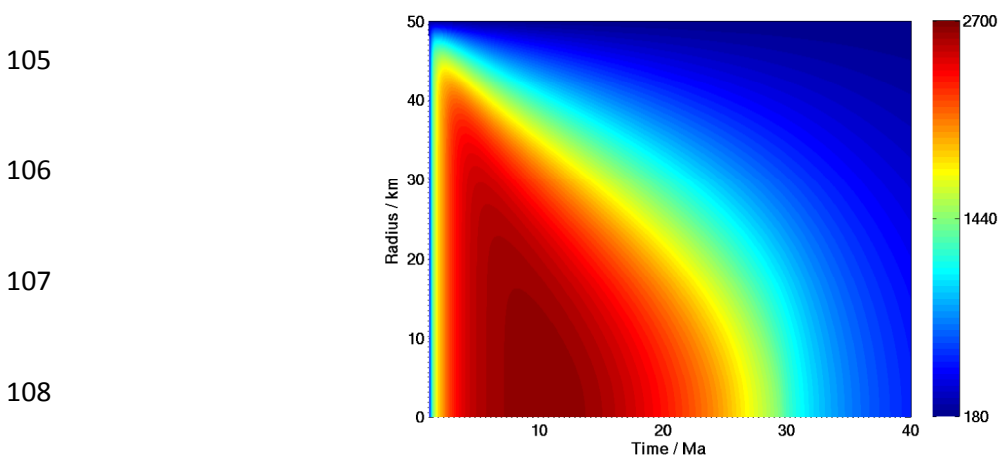

**Fig. S.3:** Temperature evolution for an R = 50 km planetesimal that accreted 1.0 Ma after CAI formation. According to Moskovitz and Gaidos (2011) their Fig. 1, the onset of melting occurs at 1.3 Ma after the formation of CAIs, at 11 Ma a peak temperature of 2615 K is

112 reached, and by 29 Ma the entire body cools to below the silicate solidus. In the simulation

113 performed with our code, onset of melting occurs at 1.34 Ma, the maximum temperature of

114 2656 K is reached at 10.65 Ma, and by 29.5 Ma the entire body cools to below the silicate

115 solidus. The difference between the maximum temperatures between Moskovitz and Gaidos

116 (2011) and our simulation is approximately 1.5 %.

117 **4. Merk et al. (2002) (melting, latent heat consumption and release, no differentiation)**

118 Parameters:

| | | | | |
|---|---|---|---|---|
| 119 | Radius | R | 5 - 105 | km |
| 120 | Formation time rel. to CAIs | $t_0$ | 0 - 1.44 | Ma |
| 121 | Thermal conductivity | k | 3.31 | W m$^{-1}$ K$^{-1}$ |
| 122 | Density | ρ | 3810 | kg m$^{-3}$ |
| 123 | Matrix specific heat capacity | $c_{p,s}$ | Debye | J kg$^{-1}$ K$^{-1}$ |
| 124 | Melt specific heat capacity | $c_{p,m}$ | 2000 | J kg$^{-1}$ K$^{-1}$ |
| 125 | Initial / surface temperature | $T_0$, $T_N$ | 290 | K |
| 126 | Porosity | φ | 0 | - |
| 127 | Initial ratio $^{26}$Al/$^{27}$Al | [$^{26}$Al/$^{27}$Al]$_0$ | 5.0x10$^{-5}$ | - |
| 128 | Initial heat production | $Q_0$ | 13.89*[$^{26}$Al/$^{27}$Al]$_0$ | W m$^{-3}$ |
| 129 | Half-life of $^{26}$Al | $t_{1/2}$ | 0.72 | Ma |
| 130 | Solidus temperature | $T_S$ | 1425 | K |
| 131 | Liquidus temperature | $T_L$ | 1850 | K |
| 132 | Latent heat of Si | L | 400 | kJ kg$^{-1}$ |

133 The treatment of consumption and release of latent heat in our model is identical to that in

134 Merk et al. (2002). However, in Merk et al. (2002) the specific heat capacity of the solid

135 matrix is obtained from the Debye law. We implemented the Debye law to be as close to this

136 model as possible. Unfortunately, the value of the Debye temperature used is not stated in

137 Merk et al. (2002). For temperatures above the solidus temperature $T_S$, the melt fraction

138 weighted average of the respective specific heats $c_{p,s}$ and $c_{p,m}$ is used. Merk et al. (2002)

139 adopted the procedure described by Miyamoto et al. (1981) for calculating $Q_0$. Here one needs

140 to account for a different density of the body, thus the value 11.67 from Miyamoto et al.

141 (1981) is scaled with the density ratio of 3810/3200 to obtain the value 13.89 from the upper

142 table.

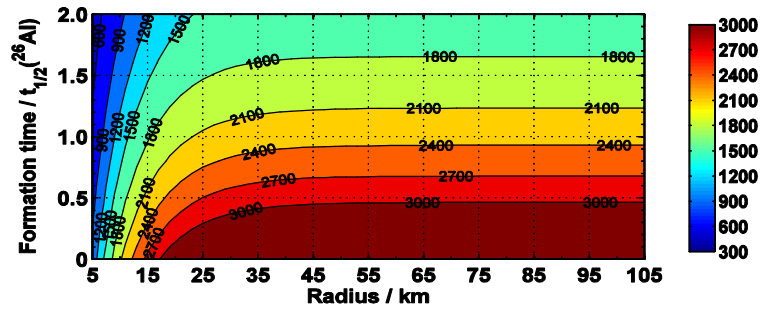

**Fig. S.4:** Maximum central temperature in planetesimals with the radii 5 – 105 km which accrete instantaneously at formation times $t_0 \leq 2*t_{1/2}(^{26}Al)$. There are minor shifts of the contour lines by less than 0.1 Ma on this plot compared to Fig. 1 in Merk et al. (2002), probably due to the unknown chosen value of the Debye temperature. Also, the lines in Merk et al. (2002) are broken due to the coarse resolution along the x-axis.

**5 Šrámek et al. (2012) (accretion and differentiation)**

Additional parameters apart from those given in section A.1.2.

| Initial radius | $R_0$ | 5 | km |
|---|---|---|---|
| Final radius | $R$ | 500 | km |
| Accretion onset time rel. to CAIs | $t_0$ | 0 | Ma |
| Accretion duration | $t_a$ | 3 | Ma |
| Accretion law | - | linear | - |
| Fe melting temperature | $T_{Fe}$ | 1261 | K |
| Si melting temperature | $T_{Si}$ | 1408 | K |
| Fe melt viscosity | $\eta_{Fe}$ | 1 | Pa s |
| Latent heat of Fe | $L_{Fe}$ | 250 | kJ kg$^{-1}$ |
| Latent heat of Si | $L_{Si}$ | 500 | kJ kg$^{-1}$ |

It should be noted that the approach on the melting process and on the consumption and release of latent heat in Šrámek et al. (2012) is completely different from ours. Also the differentiation process is modeled in another way within the framework of a two-phase flow model (although the basic idea of the density contrast as the driving force of melt percolation is the same). The goal of this comparison is to clarify, whether, in spite of all the differences, the intermediate and final structures produced by the differentiation process correspond to each other approximately, even in a rather complex case where a body accretes and differentiates simultaneously.

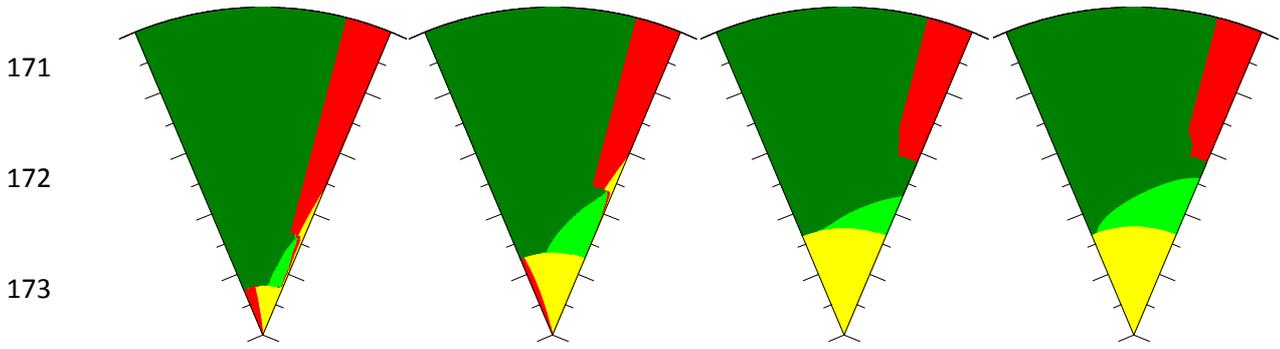

**Fig. S.5:** Volume fractions of solid metal (red), liquid metal (yellow), solid silicate (dark green), liquid silicate (light green) of the body described above at different times during and after its accretion (from left to right the times are: t = 0.4, 0.6, 3.0 and 20.0 Ma). Although the actual radius of the body increases, normalised radius is used for clarity. In our simulation a slightly smaller core and a thicker undifferentiated layer was obtained compared to Šrámek et al. (2012), Fig. 7.

**Numerical issues**

**1 Time discretisation**

We utilise the minimum of two different time steps (associated to heat conduction and melt percolation) $\Delta t \leq \min(\Delta t_c, \Delta t_m)$. Here, $\Delta t_c \leq 0.5 \frac{\rho c_p}{k}\left(1 + x_{Fe} Ste_{Fe} + x_{Si} Ste_{Si}\right)(\Delta x)^2$, where k is either the local thermal conductivity, or the effective thermal conductivity (if convection takes place) and $\Delta t_m \leq \min\left(\frac{\Delta x}{\Delta v_{Fe}}, \frac{\Delta x}{\Delta v_{Si}}\right)$. Both time steps are computed locally for each grid point of the radial discretisation, due to the variations in the spatial resolution and in the material properties. Subsequently the minimum over all grid points is adopted as the actual time step. Also, if during the convection in the core the heat flux at the CMB falls below the adiabatic one, the time step is not increased to the full value $\Delta t_c$ for the non-effective conductivity, but kept at the last value where $k_{eff}$ was involved, until convection proceeds.

**2 Spatial resolution**

The spatial resolution in the models presented in the study was set to 490 m in the interior and 160 m in the outer 25 radius% of the body (for the compacted reference). The fine resolution of the shallow regions is due to the extra-thin structures and sharp gradients in the temperature and porosity. If the resolution of the crustal region is too rough, the cooling and crystallisation of the melt in the shallow regions on its way to the surface is underestimated. Thus, eruptions occur where there should be none, and a basaltic crust forms, although the surface should remain primordial. On the other hand, if the resolution is too fine (e.g. 10 m), then no eruptions and basaltic crust formation will occur at all by porous flow, because of the effective cooling at the depth of 10 m. This is in principal possible; however, melt migration to the surface via dykes and cracks through such a thin layer is most likely. Because dykes have not been implemented in our model, the specific choice of resolution of 160 m in the outer 25 radius% is provides a compromise which allows the consideration of both the cooling and crystallisation of the ascending melt and to some extent of the melt migration via dykes. According to our test runs, a more coarse resolution increases the melt percolation rate to the surface and decreases the life-time of the shallow magma ocean, and vice versa.